\documentclass[a4paper,twocolumn,longbibliography,accepted=2026-01-15]{quantumarticle}
\pdfoutput=1
\usepackage{amsthm,amsfonts,amsmath,amssymb}
\usepackage[utf8]{inputenc}
\usepackage[english]{babel}
\usepackage[T1]{fontenc}
\usepackage[colorlinks]{hyperref}
\usepackage[normalem]{ulem}
\usepackage{soul}
\usepackage{comment}
\usepackage{graphicx}
\usepackage{braket}
\usepackage{xcolor}
\usepackage{multirow}
\usepackage[percent]{overpic}
\usepackage{accents}
\usepackage{mathtools}
\usepackage{bbm}
\usepackage{bm}
\usepackage{upgreek} 
\usepackage{isomath}
\usepackage{tikz}
\usepackage{lipsum}
\usepackage[sort&compress,numbers]{natbib}

\newcommand{\mbf}[1]{\mathbf{#1}}
\renewcommand{\rm}[1]{\mathrm{#1}}

\def\be{\begin{equation}}   
\def\ee{\end{equation}}

\definecolor{forestgreen}{rgb}{0.1, 0.6, 0.2}

\begin{document}

\title{Dissipation and non-thermal states in cryogenic cavities}

\author{Zeno Bacciconi}
\affiliation{SISSA — International School of Advanced Studies, via Bonomea 265, 34136 Trieste, Italy}
\email{zbaccico@sissa.it}
\author{Giulia Piccitto}
\affiliation{Dipartimento di Matematica e Informatica, Universit\`{a} di Catania, Catania, Italy}
\author{Alessandro Maria Verga}
\affiliation{Dipartimento di Fisica e Astronomia ``Ettore Majorana'', Universit\`{a} di Catania, , via S. Sofia 64, 95123 Catania, Italy}
\author{Giuseppe Falci}
\affiliation{Dipartimento di Fisica e Astronomia ``Ettore Majorana'', Universit\`{a} di Catania, , via S. Sofia 64, 95123 Catania, Italy}
\affiliation{INFN, Sezione di Catania, I-95123 Catania, Italy}
\author{Elisabetta Paladino}
\affiliation{Dipartimento di Fisica e Astronomia ``Ettore Majorana'', Universit\`{a} di Catania, , via S. Sofia 64, 95123 Catania, Italy}
\affiliation{INFN, Sezione di Catania, I-95123 Catania, Italy}
\author{Giuliano Chiriacò}
\affiliation{Dipartimento di Fisica e Astronomia ``Ettore Majorana'', Universit\`{a} di Catania, , via S. Sofia 64, 95123 Catania, Italy}
\affiliation{INFN, Sezione di Catania, I-95123 Catania, Italy}
\email{giuliano.chiriaco@dfa.unict.it}

\maketitle

\begin{abstract}
We study the properties of photons in a cryogenic cavity, made by cryo-cooled mirrors surrounded by a room temperature environment. We model such a system as a multimode cavity coupled to two thermal reservoirs at different temperatures. Using a Lindblad master equation approach, we derive the photon distribution and the statistical properties of the cavity modes, finding an overall non-thermal state described by a mode-dependent effective temperature. We also calculate the dissipation rates arising from the interaction of the cavity field with the external environment and the mirrors, relating such rates to measurable macroscopic quantities. These results provide a simple theory to calculate the dissipative properties and the effective temperature of a cavity coupled to different thermal reservoirs, offering potential pathways for engineering dissipations and photon statistics in cavity settings.
\end{abstract}

\section{Introduction}

Interaction between light and matter is at the heart of many fundamental physical phenomena. Recent research in this field has begun exploring more extreme regimes, revealing new effects such as the Floquet-driven dynamics of quantum systems \cite{okaPhotovoltaicHallEffect2009,goldmanPeriodicallyDrivenQuantum2014,elseFloquetTimeCrystals2016,eckardtColloquiumAtomicQuantum2017,okaFloquetEngineeringQuantum2019,tsujiNonequilibriumSteadyState2009,mciverLightinducedAnomalousHall2020,lucchiniControllingFloquetStates2022,kitamuraFloquetTopologicalSuperconductivity2022,Macri2018:Casimir,Dalvit2011}, transient phases in light-induced phenomena \cite{fausti_light-induced_2011,mitrano_possible_2016,morrison_photoinduced_2014,chiriaco_transient_2018,hruby_metastability_2018,chiriaco_voltage-induced_2018,sentef_cavity_2018,zong_evidence_2019,kogar_light-induced_2020,chiriaco_negative_2020,sun_transient_2020,chiriaco_polarity_2020} and the vacuum effects arising from interaction with cavity fields.

Cavity quantum electrodynamics (cQED) provides a versatile platform for investigating light-matter interactions at the quantum level, with significant implications for quantum technologies and fundamental physics \cite{kurizki_quantum_2015,basov_electrodynamics_2011}. In particular, the study of cavity systems at cryogenic temperatures has recently gained particular attention, since many promising phenomena occur at low temperatures, where thermal fluctuations are suppressed and quantum effects become more pronounced. Indeed, in the last few years several theoretical proposals have explored the potential of cavities as tools to control and manipulate properties and phases of matter at low temperatures, including superconductivity, topological phases, transport of charge and energy, ferroelectricity and many more \cite{schachenmayer_cavity-enhanced_2015,hagenmuller_cavity-enhanced_2017,curtis_cavity_2019,kiffner_manipulating_2019,kiffner_erratum_2019,kiffner_mott_2019,hagenmuller_cavity-assisted_2018,flick_strong_2018,nataf_no-go_2010,schlawin_cavity-mediated_2019,andolina_cavity_2019,latini_cavity_2019,kena-cohen_polariton_2019,chakraborty_long-range_2021,pilar_thermodynamics_2020,li_electromagnetic_2020,andolina_theory_2020,ashida_quantum_2020,ashida_cavity_2021,dmytruk_gauge_2021,mendez_prr2020,chiriaco_critical_2022,garcia-vidal_manipulating_2021,chiocchetta_cavity-induced_2021,schlawin_cavity_2022,bacciconi_first-order_2023,dmytruk_controlling_2022,bacciconiTopologicalProtectionMajorana2024,DagRokajTopologyChiral2024,bacciconi2024theoryfractionalquantumhall,andolina2025electrostaticsinducedbreakdownintegerquantum,giannelliDetectingVirtualPhotons2024}.

Recent experimental studies have also investigated these phenomena, including the effects of cavity on chemical reactions and topological states \cite{feist_extraordinary_2015,orgiu_conductivity_2015-2,zhong_energy_2017-1,feist_polaritonic_2017,thomas_tilting_2019,rozenman_long-range_2017,lather_cavity_2019,appugliese_breakdown_2022,graziotto2025cavityqedcontrolquantum}. In particular, Ref. \cite{jarc_cavity-mediated_2023} studied the effect of a cavity on 1T-TaS$_2$, revealing a large renormalization of its metal-insulator transition temperature, indirectly measured from the temperature of a nearby cryostat. The experimental observations can be interpreted in terms of a cavity-induced modification of thermal dissipations \cite{jarc_cavity-mediated_2023,fassioli_controlling_2024,chiriacoThermalPurcellEffect2024}, rather than microscopic effects arising from the coupling of the material to the cavity.

Indeed, dissipation \cite{breuerTheoryOpenQuantum2007,gardinerQuantumNoiseHandbook,scullyQuantumOptics1997,paladino$mathbsf1Mathbsfitf$Noise2014} is an important phenomenon which is often neglected in most theoretical models of strongly coupled light-matter systems. Dissipative processes suppress quantum correlations, and may destroy any collective phenomena that rely on the coherent coupling of matter to photons. This is particularly important, for the cryogenic cavity setups realized so far \cite{jarc_cavity-mediated_2023,jarc_tunable_2022}, since the cavity is not completely shielded from the environment at room temperature $T_e$, while the mirrors of the cavity are cooled to a cryogenic temperature $T_m$. Thus, the field in the cavity is not in a vacuum state, as the cavity modes are populated with thermal photons leaking into the cavity. Although the effects due to thermal photons are generally expected to be small, they have been shown to produce substantial temperature corrections in certain cases \cite{jarc_cavity-mediated_2023,chiriacoThermalPurcellEffect2024,pannir-sivajothi_blackbody_2024,fassioli_controlling_2024}.

It is therefore important to characterize the properties of the cavity photons, in order to identify the regimes in which dissipation and thermal effects may be safely neglected. Previous studies have explored related scenarios, such as a single-mode cavity coupled to two baths \cite{ruting_cavity_2017} or the emission statistics of a driven cavity \cite{portugalPhotonEmissionStatistics2023}. However, no comprehensive studies on multimode cavities coupled to multiple reservoirs currently exist in the literature.

The goal of this work is to bridge this gap, and provide a simple theory to calculate thermal effects in realistic cryogenic cavities. To that end, we investigate the dissipative dynamics and thermal properties of photons in a multi-mode cavity connected to two distinct reservoirs. We find that the photons are not thermally distributed, and that their properties crucially depend on the dissipation rates into the two reservoirs. This makes it possible to control and manipulate the distribution of the photons inside the cavity. We calculate the dissipation rates from a miscroscopic model and express them in terms of measurable macroscopic quantities. We also consider the effect of non-linear interactions between the photons and verify that they do not qualitatively alter the physics of the system. These are the main results of our work.

We model the cryogenic cavity as a multimode harmonic system coupled to two thermal baths at different temperatures: the cryogenically cooled mirrors at temperature $T_m$ and the warmer external environment at temperature $T_e$. We write a Lindblad master equation to describe the dissipation into the two thermal reservoirs. While our formalism is very general and applies to different cavity setups -- e.g. Fabry-Perot cavities, nanoplasmonic cavities, split ring resonators, etc \cite{maierPlasmonicsFundamentalsApplications2007,novotnyPrinciplesNanoOptics2012} -- we focus on Fabry-Perot cavities, that most accurately describe current experimental setups \cite{jarc_cavity-mediated_2023,jarc_tunable_2022}.

We initially assume that all cavity modes are independent, with each mode coupled individually to the two reservoirs through photon creation and annihilation processes. The dissipation rates satisfy a detailed balance relation at the temperature of each respective reservoir. Unlike the typical case of a cavity coupled to a single reservoir, where the photons thermalize to the temperature of the reservoir, we consider a fundamentally different situation, where the two reservoirs maintain the photons in a non-equilibrium state. Indeed, solving the Lindblad equation shows that each photon mode obeys a thermal statistic with an effective temperature that depends on the frequency of the mode, so that the cavity modes are not thermalized. The effective temperature exhibits a crossover between two regimes: a \textit{classical} regime at low frequencies where the temperature is a weighted average of $T_e$ and $T_m$, and a \textit{quantum} regime at high frequencies, where the temperature approaches the largest between the two reservoirs' temperatures.

The effective temperature crucially depends on the ratio between the dissipation rates into the mirrors and into the environment -- this is another difference with the single reservoir case. Measurements of the cavity quality factor yield the sum of the two dissipation rates, and are not sufficient to determine them separately. To address this, we analyze a microscopic model of the coupling between the cavity modes and the mirrors or the external environment, and derive an expression for the respective dissipation rates. One of the key results of this work is to quantitatively express these rates in terms of measurable macroscopic quantities, such as conductivity, reflectivity and other properties of the mirrors.

Additionally, we investigate what happens if the cavity modes are coupled to each other by non-linear interactions. The interaction induces a partial thermalization between modes at different temperatures, smoothing out the frequency dependence of the effective temperature. However, this effect is perturbative, and does not alter the qualitative physics we observed.

The remainder of this paper is structured as follows. In Section \ref{Sec:model} we introduce the cavity setup and the model, writing down the Lindblad equation. In Section \ref{Sec:Rates} we derive the Lindblad dissipative rates from a microscopic coupling and relate them to macroscopic properties of the cavity; we carry out this analysis separately for the mirrors (Sec. \ref{Sec:RatesMirror}) and the external environment (Sec. \ref{Sec:RatesEnv}). In Section \ref{Sec:Solution} we solve the Lindblad equation and analyze the properties of the photons populating the cavity. In Section \ref{Sec:nonLin} we address non linearities in the cavity which mix different modes and investigate their effects on the effective mode temperatures. Finally, in Section \ref{Sec:Conclusions} we draw our conclusions.

\section{The model}\label{Sec:model}

\begin{figure*}
    \centering
    \setlength{\unitlength}{1cm}
    \begin{picture}(0,0)
    \put(-8.3,-0.5){a)}
    \put(-0.3,-0.5){b)}
    \end{picture} 
    \includegraphics[width=\linewidth]{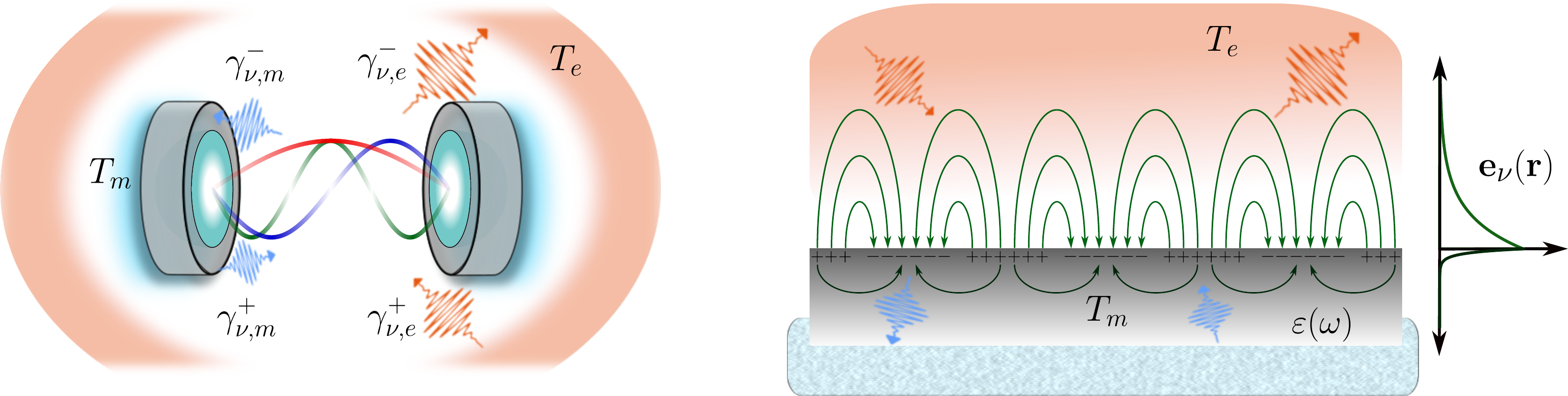}
    \caption{Sketch of the prototypical system under consideration. (a) Fabry Perot cavity with the mirrors at temperature $T_m$ and the surrounding environment at temperature $T_e$. Photons are exchanged by the cavity with the two reservoirs at the rates given in Eq. \eqref{Eq:Rates}. (b) Approximate sketch of a nanoplasmonic cavity where the field is localized near a metal-vacuum interface. Photons are again lost due to interaction with the metal -- at temperature $T_m$ and with dielectric function $\varepsilon(\omega)$ -- and with free space -- at temperature $T_e$. The plot on the left sketches the decay of the field in the metal and in the vacuum.}
    \label{fig:sketch}
\end{figure*}
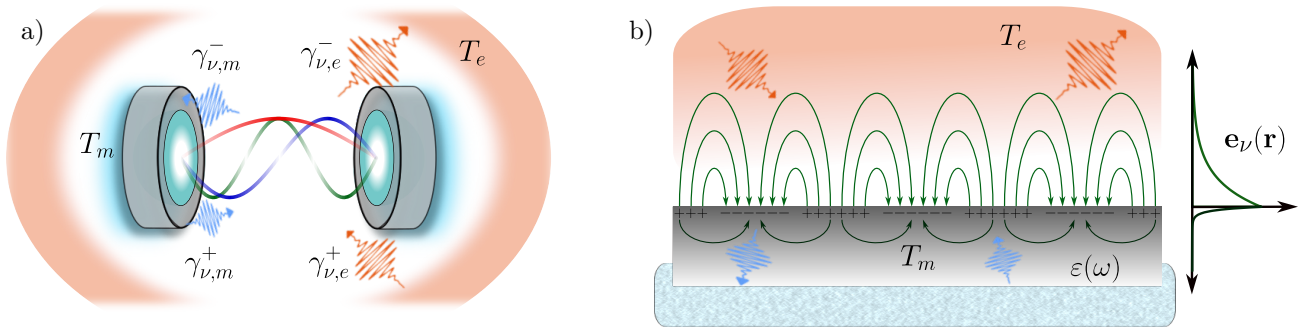

We consider a very general setup, sketched in Fig. \ref{fig:sketch}: an electromagnetic cavity that supports multiple photon modes $\nu$, each modeled as a quantum harmonic oscillator with frequency $\omega_{\nu}$ and mode volume $V_{\nu}$.

We take into account two possible dissipation mechanisms corresponding to two distinct baths for each mode $\nu$: 
\begin{itemize}
    \item emission into free space $\gamma_{\nu,e}$, i.e. the external electromagnetic environment, at temperature $T_e$\,;
    \item absorption by the mirrors $\gamma_{\nu,m}$, at a colder temperature $T_m$\,.
\end{itemize}

The presence of two distinct thermal reservoirs places the cavity in a non-equilibrium setting, characterized by structured dissipative interactions where different photon modes interact differently with each reservoir. This captures the essential dissipative dynamics of different cavity settings (Fabry-Perot, nanoplasmonic structures, photonic crystal waveguides, etc.), whose different nature is encoded in the rates $\gamma_{\nu,e/m}$. For simplicity, we will often focus on the case of Fabry-Perot cavities with cryogenically cooled mirrors \cite{jarc_cavity-mediated_2023}.

We describe the dynamics of the system using a Lindblad master equation formalism, which accounts for both unitary evolution and dissipative interactions with the reservoirs. We start from the total Hamiltonian of the system
\begin{gather}\label{Eq:Ham}
H = H_c + H_{c-m} + H_m + H_{c-e} + H_e;\\
\label{Eq:Hnu}H_c = \sum_{\nu} H_{\nu}; \qquad H_{\nu} = \omega_{\nu} a_{\nu}^\dagger a_{\nu},
\end{gather}
where $H_c$ is the cavity Hamiltonian, $H_{m/e}$ are the Hamiltonians of the mirrors and external free space respectively, $H_{c-m}$ is the coupling between cavity and mirrors, and $H_{c-e}$ is the coupling between the cavity and free space. The creation and annihilation operators for mode $\nu$ with frequency $\omega_{\nu}$ are $a_{\nu}^\dagger$ and $a_{\nu}$. 

We can derive the Lindblad master equation for the density matrix $\rho$ of the cavity field by tracing out the degrees of freedom of the mirrors and of free space. This is done explicitly in Section \ref{Sec:Rates} under the assumption of weak coupling between cavity and the two reservoirs. Nonetheless, we can already write a phenomenological expression for the Lindblad equation based on some general assumptions:
\begin{equation}\label{Eq:FullLindblad}
\frac{d\rho}{dt} = -i [H_c, \rho] + \mathcal{L}_{m}(\rho) + \mathcal{L}_{e}(\rho),
\end{equation}
where $\mathcal{L}_{m}(\rho)$ and $\mathcal{L}_{e}(\rho)$ are Lindblad superoperators representing the dissipative coupling of the cavity to the mirrors (at temperature $T_m$) and free space (at temperature $T_e$), respectively. Each mode is assumed to couple independently to the two reservoirs, with dissipation rates that are mode and reservoir-dependent, reflecting the distinct interactions with the mirrors and free space. Therefore, the Lindblad terms are written as
\begin{gather}\label{Eq:Lindbladian}
\mathcal{L}_{\alpha}(\rho) = \sum_{\nu}
\gamma_{\nu,\alpha}^-\mathcal{D}[a_{\nu}](\rho)+
\gamma_{\nu,\alpha}^+\mathcal{D}[a^{\dagger}_{\nu}](\rho);\\
\label{Eq:Dissipator} \mathcal{D}[a](\rho) = a \rho a^\dagger - \frac{1}{2} \{a^\dagger a, \rho\} ,
\end{gather}
where $\alpha = m, e$ indicates the mirror or environment reservoir, while $\gamma_{\nu,\alpha}^-$ ($\gamma_{\nu,\alpha}^+$) is the dissipation rate for the destruction (creation) of a photon from mode $\nu$ due to the interaction with reservoir $\alpha$.

The rates all satisfy the detailed balance relation given by $\gamma_{\nu,\alpha}^+=e^{-\omega_{\nu} / T_\alpha}\gamma_{\nu,\alpha}^-$. Since the exchange of photons with the reservoirs occurs via creation or destruction of bosonic excitations in the reservoirs, we can write the rates as 
\begin{gather}\label{Eq:Rates}
\gamma_{\nu,\alpha}^-=\gamma_{\nu,\alpha}(1+n_B(\omega_{\nu}, T_\alpha));\\\notag\gamma_{\nu,\alpha}^+=\gamma_{\nu,\alpha}n_B(\omega_{\nu}, T_\alpha),
\end{gather}
where $n_B(\omega_{\nu}, T_\alpha) = (e^{\omega_{\nu} / T_\alpha} - 1)^{-1}$ is the Bose-Einstein occupation number of the reservoir at temperature $T_\alpha$ and energy $\omega_{\nu}$. We have thus separated the contribution of thermal excitations in the reservoir -- appearing in the $n_B$ factor -- from that of the spectral properties of the reservoirs, which are contained in $\gamma_{\nu,\alpha}$ \footnote{In principle, $\gamma_{\nu,\alpha}$ may have a slight dependence on the temperature, due to changes with temperature in the spectrum of the reservoir. This is more likely to occur for the mirrors rather than the external free space.}. Equations \eqref{Eq:Lindbladian}-\eqref{Eq:Rates} combine to give $\mathcal{L}_{\alpha}=\sum_{\nu}\mathcal{L}_{\nu,\alpha}$ with
\begin{equation}\label{Eq:Lnualpha}
\mathcal{L}_{\nu,\alpha}=\gamma_{\nu,\alpha}\left( \left(n_B + 1 \right) \mathcal{D}[a_{\nu}](\rho) + n_B\mathcal{D}[a^{\dagger}_{\nu}](\rho) \right),
\end{equation}
where, for simplicity of notation, we omitted the dependence on $\omega_\nu$ and $T_\alpha$.

The master equation \eqref{Eq:FullLindblad}-\eqref{Eq:Lnualpha} allows us to analyze the non-equilibrium steady state of the cavity, where the cavity modes can each acquire a distinct effective temperature depending on the relative strengths of $\gamma_{\nu,m}$ and $\gamma_{\nu,e}$.

\section{Derivation of the dissipation rates}\label{Sec:Rates}

In this section we derive the phenomenological Lindblad equation~\eqref{Eq:FullLindblad}, starting from the microscopic Hamiltonian~\eqref{Eq:Ham}, for both the interaction with the mirrors and with free space.

The two derivations are treated separately in what follows. Before detailing out the calculations, we report the results, which allow to evaluate the dissipation rates in terms of macroscopic quantities.

The dissipation rate for the mirrors is given by
\begin{equation}\label{eq:rates_microscopic_mir0}
\boxed{\gamma_{\nu,m}=\frac{\sigma(\omega_\nu)}{2\epsilon_0} \frac1{V_{\nu}}\int_m d^3\mbf r|{\mbf e}_\nu(\mbf r)|^2.}
\end{equation}

Here $\sigma(\omega_{\nu})$ is the (real part of the) conductivity of the mirrors at frequency $\omega_{\nu}$, $\epsilon_0$ is the vacuum dielectric constant. The volume of the mode $\nu$ is $V_{\nu}=\int d^3\mbf r\varepsilon(\mbf r)|{\mbf e}_\nu(\mbf r)|^2$, with $\varepsilon(\mbf r)$ the dielectric constant and $\mbf{e}_{\nu}$ the mode function of the electric field; since the mode function is mostly confined inside the cavity where $\varepsilon(\mbf{r})=1$ we approximate the volume as $V_{\nu}\approx\int d^3\mbf r|{\mbf e}_\nu(\mbf r)|^2$.\footnote{This is also true for plasmonic cavities, where most of the mode volume comes from the vacuum contribution, see Appendix \ref{App:Plamonics}.} The integral $\int_m$ is calculated only over the volume of the mirrors. Equation \eqref{eq:rates_microscopic_mir0} can be interpreted as the power dissipated by Joule heating in the mirrors, divided by the energy of the cavity electromagnetic field.

The dissipation rate into free space is given by
\begin{equation}\label{eq:rates_microscopic_env0}
\boxed{\gamma_{\nu,e} = \sum_{\mathbf{k}, \lambda} \frac{\pi \omega_{\nu}^2}{4V V_{\nu}} \left| \int d^3\mathbf{r} \,\varepsilon(\mbf r) \mathbf{f}_{\mathbf{k}, \lambda}(\mathbf{r}) \cdot \mathbf{e}_{\nu}^*(\mathbf{r}) \right|^2 \delta(\omega_{\nu} - \omega_\mbf k).}
\end{equation}

Here $\mathbf{f}_{\mathbf{k}, \lambda}$ is the mode function of free space mode $\mbf k,\lambda$, and $V=\int d^3\mbf r|{\mbf f}_{\mbf k, \lambda}(\mbf r)|^2$ is its volume. Equation \eqref{eq:rates_microscopic_env0} can be effectively interpreted as the Fermi golden rule for the decay from the cavity modes into free space modes: it contains the relative overlap between the mode functions, summed over the modes with energy equal to $\omega_\nu$.

We reiterate that Eqs. \eqref{eq:rates_microscopic_mir0} and \eqref{eq:rates_microscopic_env0} are very general, as they only require the knowledge of the conductivity and dielectric properties of the mirrors, which also determine the mode functions of both the cavity and free space.

Note that, when considering the dissipation due to one reservoir, we ignore the effect of the other reservoir. This assumption is justified if the mirrors and free space are uncorrelated and their state wavefunction factorizes. This is not true in general, as the mirrors and free space can couple via an extra $H_{m-e}$ term in the Hamiltonian \eqref{Eq:Ham}. However, $H_{m-e}$ is typically very small, of the same order of magnitude as the coupling between mirrors and cavity $H_{c-m}$. Within perturbation theory, the correlations between mirrors and environment produce a small correction of order $\sim H_{m-e}^2/\omega_\nu^2$ to the dissipation rates. Therefore, we can safely ignore the coupling between mirrors and external environment for the calculations that follow. 

\subsection{Mirror-cavity interaction}\label{Sec:RatesMirror}

In this section we microscopically derive the dissipative dynamics due to the mirror-cavity interaction, i.e. $\mathcal{L}_m(\rho)$ in Eq.~\eqref{Eq:FullLindblad}. 

The interaction between cavity and mirrors is assumed to be a dipole interaction as 
\begin{equation}
H_{c-m}= - \int_m d^3\mbf r\,{\mbf d}(\mbf r) \cdot {\mbf E}(\mbf r),
\end{equation}
where the integral is performed over the mirror volume. Here ${\mbf d} = (d_x({\mbf r}), d_y ({\mbf r}), d_z({\mbf r}))$ is the \emph{dipole density} operator of the mirror, and 
\begin{equation} 
 {\mbf E}(\mbf r) = i \sum_{\nu} \sqrt{\frac{\omega_{\nu}}{2\epsilon_0V_{\nu}}} \big({\mbf e}_{\nu}(\mbf r) a_{\nu} - {\mbf e}_{\nu}^*(\mbf r) a^\dagger_{\nu}\big) 
 \label{eq:efield}
\end{equation}
the operator describing the electric field. The field polarization vector is given by ${\mbf e}_{\nu}$, with $\nu$ labelling the photonic mode, and the mode volume is defined as $V_{\nu} = \int |{\mbf e}_{\nu}({\mbf r})|^2 d{\mbf r}$.

The Lindblad master equation describing the dissipation due to this interaction is derived from the microscopic model \cite{breuerTheoryOpenQuantum2007,dabbruzzoSelfconsistentMicroscopicDerivation2021}
\begin{equation}\label{eq:L_m}
	\begin{aligned}
		\mathcal{L}_\text{m}&[\rho] = -\sum_{\omega} \int d^3 \mbf r d^3 \mbf r' \Gamma(\mbf  r- \mbf r', \omega) \cdot\\
         &\cdot\Big[E_\omega(\mbf r) \rho E^\dagger_\omega(\mbf r')-\frac{1}{2}\left \{ E_\omega^\dagger(\mbf r') E_\omega(\mbf r), \rho \right\}\Big].
  \end{aligned}
\end{equation}

Here  
\begin{equation}
    \Gamma(\mbf r, \omega)\equiv \int_0^{\infty}d\tau e^{i\omega\tau}\braket{\mbf d(\mbf r,\tau)\mbf d(0,0)}
\end{equation}
is the one-side Fourier transform of the two-times dipole correlation function, and ${\mbf E}_\omega(\mbf r)={\mbf E}_\omega^\text{em}(\mbf r)+{\mbf E}_\omega^\text{ab}(\mbf r)$ is the electric field operator expressed in the energy eigenbasis (for the derivation see App.~\ref{app:eigenoperators}), where
\begin{equation}\label{Eq:em_ab}
\begin{aligned} &{\mbf E}_\omega^\text{em}(\mbf r) = i\sum_\nu\sqrt{\frac{\omega_{\nu}}{2\epsilon_0V_{\nu}}} \delta(\omega_\nu - \omega){\mbf e}_\nu ({\mbf r})\,a_\nu,\quad\\
&{\mbf E}_\omega^\text{ab}(\mbf r) = -i\sum_\nu\sqrt{\frac{\omega_{\nu}}{2\epsilon_0V_{\nu}}} \delta(\omega_\nu + \omega){\mbf e}_\nu^* ({\mbf r})\,a_\nu^\dagger,
\end{aligned}
\end{equation}
identify, for any frequency $\omega$, two jump operators. We notice that $E_\omega^\text{ab}E_\omega^\text{em}=0$, and thus we can write Eq.~\eqref{eq:L_m} as the sum of two contributions
\begin{equation}
		\mathcal{L}_\text{m}[\rho]=\mathcal{L}_\text{m}^\text{em}[\rho] + \mathcal{L}_\text{m}^\text{ab}[\rho],
\end{equation}
with
\begin{gather}
		\mathcal{L}_\text{m}^\text{em}[\rho] = \sum_\nu\gamma_{m,\nu}^- \Big[a_\nu \rho a^\dagger_\nu - \frac{1}{2} \big\{ a^\dagger_\nu a_\nu, \rho \big\}\Big];\\
		\mathcal{L}_\text{m}^\text{ab}[\rho] = \sum_\nu\gamma_{m,\nu}^+ \Big[a^\dagger_\nu \rho a_\nu - \frac{1}{2} \big\{ a_\nu a^\dagger_\nu, \rho \big\}\Big],
\end{gather}
\begin{equation}
		\gamma_{m,\nu}^{\pm}=\frac{\omega_\nu}{2\epsilon_0 V_{\nu}}\, \int d^3\mbf r\,d^3\mbf r'\Gamma(\mbf r- \mbf r',\mp\omega_{\nu}){\mbf e}_\nu(\mbf r){\mbf e}_\nu^*(\mbf r').
\end{equation}

The spatial integral can be rewritten as $\int d^3\mbf q\tilde\Gamma(\mbf q, \omega_{\nu})|\tilde{\mbf e}_\nu(\mbf q)|^2$ by Fourier transforming to momentum space where $\tilde\Gamma(\mbf q, \omega)=\int d^3\mbf re^{-i\mbf q\cdot\mbf r}\Gamma(\mbf r, \omega)$. Since the electric field typically changes over distances much larger than the microscopic lengthscales of the mirrors, $\tilde\Gamma$ is approximately constant for small $\mbf q$ and we can write
\begin{equation}
		\gamma_{m,\nu}^{\pm}\approx\frac{\omega_\nu}{2\epsilon_0 V_{\nu}}\, \tilde\Gamma(0,\mp\omega_{\nu})\int d^3\mbf r|{\mbf e}_\nu(\mbf r)|^2
\end{equation}

Using the Lehmann representation, see Appendix \ref{app:diss_conductivity}, it is possible to show that $\Gamma(-\mbf r,-\omega)=e^{-\beta_m\omega}\Gamma(\mbf r,\omega)$ -- where $\beta_m=1/T_m$ is the inverse temperature of the mirrors -- so that $\tilde\Gamma(0,-\omega_{\nu})=e^{-\beta_m\omega_{\nu}}\tilde\Gamma(0,\omega_{\nu})$.

We can also relate $\Gamma$ to the electric conductivity $\sigma$ via the fluctuation-dissipation theorem. We use the Kubo formula \cite{Kubo}
\begin{equation}
	\sigma(\mbf q,\,\omega) = \frac{1}{i} \int_0^{\infty} dt e^{i\omega t} \int d^3\mbf re^{-i\mbf q\cdot\mbf r}\braket{[{\mbf d}(\mbf r,t), {\mbf J}]},
\end{equation}
where the electric current density is ${\mbf J}(\mbf r,t)=\partial_t{\mbf d}(\mbf r,t)$. Deriving with respect to time inside the integral and integrating by parts, we find $\sigma(\mbf q,\,\omega) = \omega\int_0^{\infty} dt e^{i\omega t} \int d^3\mbf re^{-i\mbf q\cdot\mbf r}\braket{[{\mbf d}(\mbf r,t), {\mbf d}]}$. Explicitly writing out the expectation values in the Lehman representation (Appendix \ref{app:diss_conductivity}) we obtain
\begin{equation}
	\begin{aligned}
		\sigma(\mbf q, \omega) =\omega \,(1-e^{-\beta_m\omega})\,\tilde\Gamma ( \mbf q, \omega).
	\end{aligned}
\end{equation}

This leads to the final expression for the jump rates
\begin{gather}
\notag
\gamma_{m,\nu}^{\pm}=\pm\frac{1}{2\epsilon_0V_{\nu}}  \frac{  \sigma(\omega_\nu)}{e^{\pm\beta_m\omega_\nu}-1}	\int d^3\mbf r|{\mbf e}_\nu(\mbf r)|^2;\\
\label{eq:rates_microscopic}
\gamma_{m,\nu}=\frac{\sigma(\omega_\nu)}{2\epsilon_0V_{\nu}}\int_m d^3\mbf r|{\mbf e}_\nu(\mbf r)|^2,
\end{gather}
where we have reintroduced the label $m$ for the integral performed over the mirror space.

\begin{figure}
    \centering
    \includegraphics[width=\columnwidth]{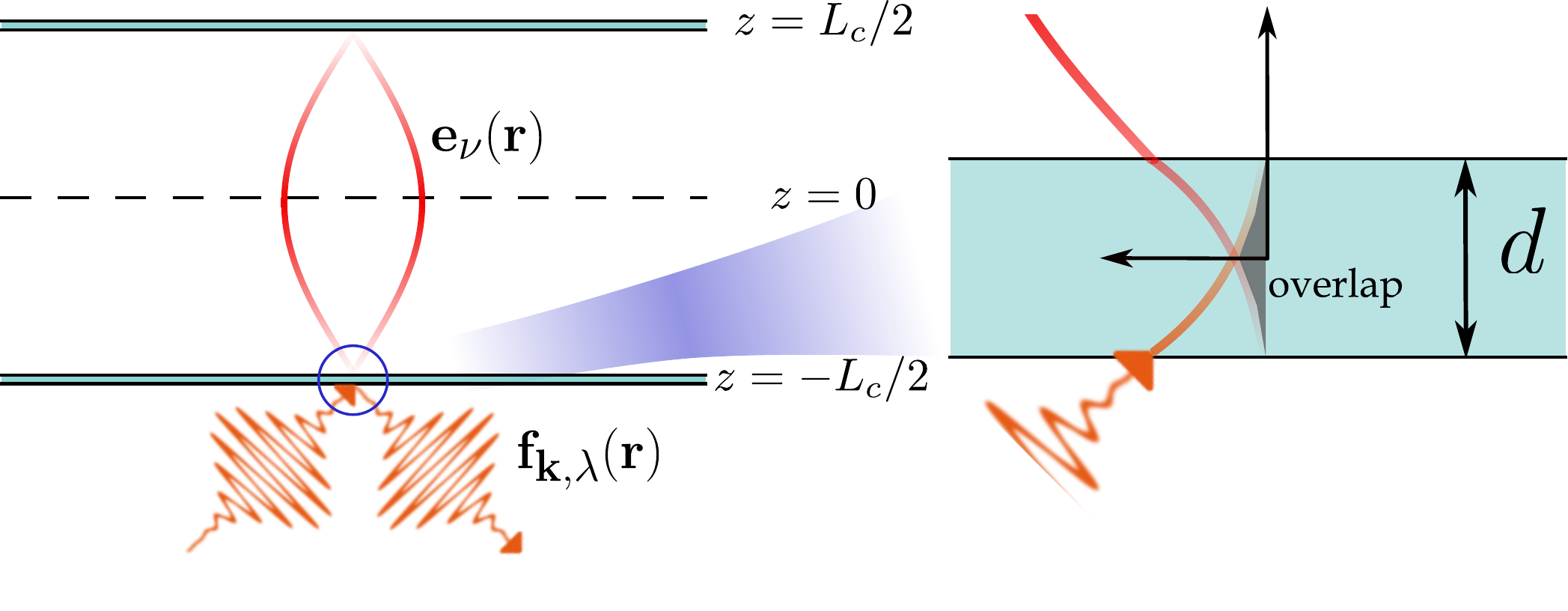}
    \caption{Skectch of the electric field inside and outside of a Fabry-Perot cavity. The cavity is formed by two mirrors of thickness $d$ located at $z=\pm L_c/2$. The fields modes inside the cavity are represented by $\mbf e_\nu(\mbf r)$ while the free space modes are $\mbf f_{\mbf k,\lambda}(\mbf r)$. The zoomed panel shows the overlap between the cavity modes and the free space modes, both of which decay exponentially inside the mirrors.}
    \label{fig:overlapFields}
\end{figure}

\subsection{Environment dissipation rate}\label{Sec:RatesEnv}

In this subsection we consider the coupling between the free space modes and the cavity modes and calculate the related dissipation rate. This coupling arises from the overlap between the cavity fields and the free space fields, which occurs inside the mirrors, see Fig. \ref{fig:overlapFields}.

The electric field in space can be written as the sum of two terms $\mbf E(\mbf r) = \mbf E_c(\mbf r) + \mbf E_e(\mbf r)$, where $\mbf E_c(\mbf r)$ is the cavity field as in Eq. \eqref{eq:efield} and
\begin{equation}\label{Eq:EfieldFree}
\mbf{E}_e(\mbf{r}, t) = i \sum_{\mbf{k}, \lambda} \sqrt{\frac{\omega_k}{2 \epsilon_0 V}}  a_{\mbf{k}, \lambda} \, \mbf{f}_{\mbf{k}, \lambda}(\mbf{r}) e^{-i \omega_\mbf k t}-\text{h.c.},
\end{equation}
is the free space field. The modes are labeled by the wavevector $\mbf k$ and by $\lambda$, which accounts for polarization and for other quantum numbers that may arise from the geometry of the system. The frequency of the mode is $\omega_\mbf k$ and $\mbf f_{\mbf k,\lambda}$ is the respective mode function. In general, the environment modes $\mbf f_{\mbf k,\lambda}$ vanish inside the cavity, while the cavity modes $\mbf e_{\nu}$ vanish in the space outside the cavity, so that they overlap only inside the mirror, as discussed more in detail in the next subsection.

The total energy of the system is written as $\frac{\epsilon_0}{2}\int d\mbf r\varepsilon(\mbf r)\mbf E(\mbf r)^2$, where $\varepsilon(\mbf r)$ is the dielectric function at position $\mbf r$. The magnetic field gives a contribution equal to that of the electric field, while the time average over the harmonic oscillations of the field contributes a factor $1/2$. Taking the square of $\mbf E$, we obtain $\emph{(i)}$ a term $\mbf E_c^2$ that gives the Hamiltonian $H_c$ of the cavity after quantization; $\emph{(ii)}$ a term $\mbf E_e^2$ that gives the Hamiltonian $H_e=\sum_{\mbf k,\lambda}\omega_{\mbf k}a_{\mbf{k}, \lambda}^{\dagger} a_{\mbf{k}, \lambda}$ of the free space modes; \emph{(iii)} a cross term $2\mbf{E}_e(\mbf r)\cdot\mbf E_c(\mbf r)$ which represents the coupling between cavity and free space.

Therefore, we can write the coupling Hamiltonian as
\begin{equation}\label{Eq:Hce}
H_{c-e}=\epsilon_0\int_m d^3\mbf r\varepsilon(\mbf r)\mbf{E}_e(\mbf r)\cdot\mbf E_c(\mbf r),
\end{equation}
where the subscript $m$ indicates that the support of the integral is given by the mirrors, where the cavity and free space modes overlap.

In order to calculate the dissipation rate we proceed exactly as in Section \ref{Sec:RatesEnv} with the substitution $-\mbf d\rightarrow\epsilon_0\mbf E_e$. We obtain
\begin{equation}\label{Eq:gammaEnvpm}
\gamma_{\nu,e}^{\pm}=\frac{\epsilon_0 }2\frac{\omega_{\nu}}{V_{\nu}}\, \int_m d^3\mbf r\,d^3\mbf r'\Gamma_e(\mbf r, \mbf r',\mp\omega_{\nu})\tilde{\mbf e}^*_\nu(\mbf r)\tilde{\mbf e}_\nu(\mbf r')
\end{equation}
where $\tilde{\mbf e}_\nu(\mbf r')=\varepsilon(\mbf r){\mbf e}_\nu(\mbf r')$  and
\begin{equation}\label{Eq:GammaEnv}
\Gamma_e(\mbf r,\mbf r ',\omega)=\int_0^{\infty}dt e^{i\omega \tau}\langle \mbf{E}_e(\mbf{r}, \tau) \mbf{E}_e(\mbf{r}', 0) \rangle.
\end{equation}

We use Eq. \eqref{Eq:EfieldFree} to calculate the expectation value $\langle \mbf{E}_e(\mbf{r}, \tau) \mbf{E}_e(\mbf{r}', 0) \rangle$ :

\begin{gather*}
\langle \mbf{E}_e(\mbf{r}, \tau) \mbf{E}_e(\mbf{r}', 0) \rangle = \sum_{\mbf{k}, \lambda,\mbf k',\lambda'} \frac{\sqrt{\omega_\mbf k\omega_{\mbf k'}}}{2 \epsilon_0 V} \mbf{f}_{\mbf{k}, \lambda}(\mbf{r}) \mbf{f}_{\mbf{k}', \lambda'}^*(\mbf{r}')\cdot\\
\cdot\Big( \langle a_{\mbf{k}, \lambda} a_{\mbf{k}', \lambda'}^\dagger \rangle e^{-i \omega_\mbf k \tau} + \langle a_{\mbf{k}, \lambda}^\dagger a_{\mbf{k}', \lambda'} \rangle e^{i \omega_\mbf k \tau} \Big).
\end{gather*}

For a thermal state at temperature $T_e$, the expectation values are diagonal in $\mbf k$ and $\lambda$, and involve the thermal distribution $n_{\mathbf{k}} = \frac{1}{e^{ \omega_\mbf k /T_e} - 1}$:

\begin{gather*}
\langle \mathbf{E}_e(\mathbf{r}, \tau) \mathbf{E}_e(\mathbf{r}', 0) \rangle = \sum_{\mathbf{k}, \lambda} \frac{\omega_k}{2 \epsilon_0 V} \mathbf{f}_{\mathbf{k}, \lambda}^*(\mathbf{r}) \mathbf{f}_{\mathbf{k}, \lambda}(\mathbf{r}') \cdot\\
\cdot\Big( (1 + n_{\mathbf{k}}) \, e^{-i \omega_k \tau} + n_{\mathbf{k}} \, e^{i \omega_k \tau} \Big).
\end{gather*}

Substituting into Eq. \eqref{Eq:GammaEnv}, we perform the Fourier transform ($\int_0^{\infty}dte^{i\omega t})=\pi\delta(\omega)-i\mathcal{P}\frac1\omega$ and retain the real part, obtaining

\begin{equation}\label{Eq:Gammanu}
\Gamma_e(\mbf r,\mbf r ',\omega_\nu)=\sum_{\mathbf{k}, \lambda} \frac{\pi\omega_\mbf k}{2 \epsilon_0 V} (1 + n_{\mathbf{k}}) \, \mathbf{f}_{\mathbf{k}, \lambda}(\mathbf{r}) \mathbf{f}_{\mathbf{k}, \lambda}^*(\mathbf{r}') \delta(\omega_\nu-\omega_\mbf k)
\end{equation}
where we discarded the second term since it yields a $\delta(\omega_\nu+\omega_\mbf k)$ factor which is always zero given the positivity of both energies. Combining with Eq. \eqref{Eq:gammaEnvpm} we find
\begin{equation}\label{Eq:gamma_env}
    \begin{split}
        &\gamma_{\nu,e} = \sum_{\mathbf{k}, \lambda} \frac{\pi \omega_{\nu}^2}{4V V_{\nu}} \left| \mathcal{A}_{\mbf k,\lambda,\nu} \right|^2 \delta(\omega_{\nu} - \omega_\mbf k);\\
& \mathcal{A}_{\mbf k,\lambda,\nu}\equiv\int_m d^3\mathbf{r} \,\varepsilon^*(\mbf r) \mathbf{f}_{\mathbf{k}, \lambda}(\mathbf{r}) \cdot \mathbf{e}_{\nu}^*(\mathbf{r}).
    \end{split}
\end{equation}

This expression contains $\mathcal{A}_{\mbf k,\lambda,\nu}$, the relative overlap between the cavity mode functions \(\mathbf{e}_{\nu}(\mathbf{r})\) and the free-space mode functions \(\mathbf{f}_{\mathbf{k}, \lambda}(\mathbf{r})\) weighted by the dielectric function, summed over all free space modes at energy $\omega_{\nu}$.

\subsection{Discussion and examples}\label{Sec:RatesExample}

The expressions for the dissipation rates Eqs. \eqref{eq:rates_microscopic_mir0} and \eqref{eq:rates_microscopic_env0} are quite simple, intuitive and require the knowledge of a few ingredients.

The first two ingredients are the conductivity $\sigma(\omega)$ and the dielectric constant $\varepsilon(\omega)$ of the mirrors (or any other cavity medium). These are easily obtained experimentally via transport or spectroscopic measurements.

The other ingredients are the mode functions of the cavity $\mbf e_\nu$ and of the external environment $\mbf f_{\mbf k,\lambda}$. These are obtained by solving the Maxwell equations -- which reduce to the Helmoltz equation for the electric field -- for the geometry and the dielectric properties of the cavity. Such solution can be obtained analytically in some cases (e.g. a Fabry-Perot cavity) or numerically for more complicated cavity structures.

This point requires a more detailed discussion. In free space, the Helmoltz equation can be written as an eigenvalue problem of a self-adjoint differential operator. This way, the modes of the electromagnetic field are orthonormal, leading to a vanishing integral in Eq. \eqref{eq:rates_microscopic_env0}. However, this argument is not entirely accurate when dielectric media (or boundary conditions) -- which naturally divide the space into a cavity region and an external free space region \cite{viviescasFieldQuantizationOpen2003} -- are introduced. In such case, it is still possible to find orthonormal modes that separately satisfy the Helmoltz equation in each region (e.g. inside and outside the Fabry-Perot cavity, or the metallic substrate and the region above it for nanoplasmonic cavities) but at the price of some extra terms in the set of differential equations, which couple the modes in the two regions. These terms lead to an overlap between $\mbf e_\nu$ and $\mbf f_{\mbf k,\lambda}$, which typically occur at (or near) the boundary between the cavity region and the free space region. For a more formal discussion of these aspects, see for example Ref. \cite{viviescasFieldQuantizationOpen2003}.

In the following, we estimate the dissipation rates for two specific cavity setups: a Fabry-Perot cavity and a nanoplasmonic cavity.

\emph{Fabry Perot cavity} -- We consider a cavity delimited by two planar mirrors located at $z=\pm L_c/2$, so that fundamental frequency is  $\omega_c=\pi c/L_c$. The mirrors have thickness $d\ll L_c$ and conductivity $\sigma(\omega)$, see Fig. \ref{fig:overlapFields}. We assume the mirrors to be made of very good metals and to be highly reflective, so that the cavity has a high quality factor. Thus, in the range of frequencies of interest for our case, the negative dielectric constant is $\varepsilon(\omega)=1-\omega_p^2/\omega^2$, with the plasma frequency satisfying $\omega_p\gg\omega$. In such scenario, the electromagnetic field is exponentially suppressed inside the mirrors, with decay length $l_d\sim c/\omega_p$. The assumption of a high quality factor is equivalent to requiring that the mirrors are thick enough compared to the decay length, i.e. $\omega_pd/c\gg1$. 

As detailed in Appendix \ref{App:FPcavity}, in this regime the Fresnel equations dictate that the amplitude of the field inside the mirrors is reduced by a factor $\omega/\omega_p$ compared to the field in the cavity volume. Therefore, we write the ratio $\frac1{V_{\nu}}\int_m d^3\mbf r|{\mbf e}_\nu(\mbf r)|^2$ in Eq.~\eqref{eq:rates_microscopic_mir0} as the square of the amplitudes ratio times the ratio between the decay length and cavity length  $\sim \left(\frac{\omega_\nu}{\omega_p}\right)^2 \times \frac{l_d}{L_c}$. Finally, we substitute the expression of $\omega_c$ and write the dissipation rates as
\begin{equation}\label{Eq:gamma_mir_FPC}
\gamma_{\nu,m}\sim\frac{\sigma(\omega_\nu)}{\epsilon_0}\left(\frac{\omega_\nu}{\omega_p}\right)^2\frac{\omega_c}{\omega_p}\sim\omega_\nu\frac12\frac{\sigma(\omega_\nu)}{\epsilon_0\omega_p}\frac{\omega_\nu\omega_c}{\omega_p^2}
\end{equation}

A similar estimation can be performed to evaluate the overlap $\mathcal{A}$ of cavity modes and free space modes inside the mirrors in Eq. \eqref{Eq:gamma_env}. Again from Fresnel equations, we find that both fields are suppressed by a factor $\omega/\omega_p$ inside the mirror and exhibit an exponential decay $e^{-z\omega_p/c}$ ($e^{-(d-z)\omega_p/c}$) away from the surface that separates the mirror from free space (from the cavity). Therefore, their overlap integral $\mathcal{A}$ is simply given by $\sim (\omega/\omega_p)^2de^{-\omega_pd/c}\varepsilon(\omega)\sim de^{-\omega_pd/c}$, where we have used $\varepsilon(\omega)\sim\omega_p^2/\omega^2$. Since the cavity mode volume is $\sim L_c\sim1/\omega_c$, we estimate the dissipation rate as
\begin{equation}\label{Eq:gamma_env_FPC}
\gamma_{\nu,e}\sim\omega_\nu^2\omega_c\frac{d^2}{c^2}e^{-2\omega_pd/c}
\end{equation}
Combining Eqs. \eqref{Eq:gamma_mir_FPC} and \eqref{Eq:gamma_env_FPC} we observe that the ratio of dissipation rates is given by:
\begin{equation}
    \gamma_{\nu,m}/\gamma_{\nu,e}\sim\sigma(\omega_\nu)\;,
\end{equation} 
i.e. all the frequency dependence of the ratio of dissipation rates arise from the conductivity of the mirrors. Since most Fabry-Perot cavities operate in a frequency regime where the mirror conductivity is constant and equal to its dc value, they are well described by a frequency-independent $\gamma_m/\gamma_e$.

\emph{Nanoplasmonic cavity} -- We consider a very simple model of a nanoplasmonic cavity. A metallic substrate occupies the half space at $z<0$, with vacuum at $z>0$. The metal has dielectric constant $\varepsilon(\omega)=1-\omega_p^2/\omega^2$ and conductivity $\sigma(\omega)$, where $\omega_p\gg\omega$ is the plasma frequency. This configuration supports localized plasmonic modes that propagate along the $x$ direction and decay exponentially away from the metallic surface in the $z$ direction. The metallic substrate acts as the confining element of the cavity, similarly to the mirrors of a Fabry-Perot cavity.

The solution of the Maxwell equations, see Appendix \ref{App:Plamonics}, tells us that the electric field of a plasmon with frequency $\omega$ decays over a distance $c/\omega_p$ inside the metal and over a (much larger) distance $c\omega_p/\omega^2$ in the vacuum \cite{novotnyPrinciplesNanoOptics2012}. Moreover, the magnitude of the plasmonic field within the metal is smaller than the magnitude of the field in vacuum by a factor $\omega/\omega_p$. Thus, the main contribution to the volume of the mode $V_\nu\sim\int|\mathbf{e}_{\nu}|^2\sim c\omega_p/\omega^2$ comes from the field in the vacuum.

The dissipation rate into the metal is then roughly equal to $\sim\frac1{V_\nu}\int_m|\mathbf{e}_{\nu}|^2\sim1/(c\omega_p/\omega^2)\times(\omega/\omega_p)^2\times(c/\omega_p)$, where the first factor is the ratio of the square of the fields amplitude and the second factor is the ratio of the penetration lengths in the metal and in the vacuum. Precise calculations yield
\begin{equation}\label{Eq:gamma_mir_pla}
\gamma_{\omega,m}=\frac{\sigma(\omega)}{2\epsilon_0}\left(\frac{\omega}{\omega_p}\right)^4=\omega\frac12\frac{\sigma(\omega)}{\epsilon_0\omega_p}\left(\frac{\omega}{\omega_p}\right)^3.
\end{equation}

The calculation of the dissipation rate into free space is more complicated, because the free space modes and the plasmonic modes overlap both in vacuum and inside the metal. The leading contribution arises from the overlap of the modes in the vacuum. We have to consider a finite propagation length of the plasmon along the metallic surface -- which originates from losses in the metal -- in order to break translational invariance, and thus momentum conservation, which would otherwise forbid any coupling between plasmons and free space modes \cite{novotnyPrinciplesNanoOptics2012}. Carrying out the calculations, see Appendix \ref{App:Plamonics}, we find
\begin{equation}\label{Eq:gamma_env_pla}
\gamma_{\omega,e}\sim\omega\left(\frac{\sigma(\omega)}{\epsilon_0\omega}\right)^3\left(\frac{\omega}{\omega_p}\right)^4\sim\omega\frac{\omega}{\omega_p}\left(\frac{\sigma(\omega)}{\epsilon_0\omega_p}\right)^3.
\end{equation}
In the case of nanoplasmonic cavities the ratio between the dissipation rates is then:
\begin{equation}
    \gamma_{\omega,m}/\gamma_{\omega,e}\sim(\omega\epsilon_0/\sigma(\omega))^2\;.
\end{equation}
Assuming a typical Drude-Lorentz model, the conductivity is constant at low frequency and decays as $\sigma(\omega)\sim1/\omega^2$ at high frequencies. Thus, $\gamma_{\omega,m}/\gamma_{\omega,e}\sim\omega^2$ and environment dissipation dominates at low frequencies, while $\gamma_{\omega,m}/\gamma_{\omega,e}\sim\omega^6$ and dissipation into the metal dominates at high frequencies.

\section{Non-thermal cavity population}\label{Sec:Solution}

In this section we proceed to solve the Lindblad equation and find its steady state solution.

Under the assumptions of Section \ref{Sec:model}, each mode is independent, and the density matrix of the photon steady state factorizes into separate subspaces for each mode.\footnote{This is true even independently of the initial state, since inter-mode correlations decay exponentially over time.} Thus $\rho = \bigotimes_{\nu} \rho_{\nu}$, where $\rho_{\nu}$ is the density matrix for mode $\nu$, and the Lindblad equation for the entire system simplifies to a set of independent Lindblad equations for each mode \cite{breuerTheoryOpenQuantum2007,gardinerQuantumNoiseHandbook,scullyQuantumOptics1997} according to:

\begin{equation}\label{Eq:limblad_1mode}
\frac{d \rho_{\nu}}{dt} = -i [H_{\nu}, \rho_{\nu}] + \mathcal{L}_{\nu,m}(\rho_{\nu}) + \mathcal{L}_{\nu,e}(\rho_{\nu}),
\end{equation}
where $\mathcal{L}_{\nu,m}$ and $\mathcal{L}_{\nu,e}$ are defined in Eq. \eqref{Eq:Lnualpha}.

We notice that Eq. \eqref{Eq:limblad_1mode} is equivalent to the Lindblad equation describing the coupling to a single thermal bath with total loss and gain rates:
\begin{align}\Gamma^+_{\nu}=\gamma_{\nu,e}^++\gamma^+_{\nu,m}\qquad
    \Gamma^-_{\nu}=\gamma_{\nu,e}^-+\gamma^-_{\nu,m} \;.
\end{align}
These satisfy a detailed balance condition for an effective temperature $T^{*}_\nu$:
\begin{align}\label{eq:detailbal_Teff}
    \frac{\Gamma^+_{\nu}}{\Gamma^-_{\nu}}=e^{-\omega_\nu/T^{*}_\nu}\;.
\end{align}
Rearranging this detailed balance condition we find the effective temperature of the mode $\nu$ as:
\begin{align}\label{eq:Teff}
    \boxed{T^{*}_\nu= \frac{\omega_\nu}{-\log\left(r_{\nu,e}^- e^{-\omega_\nu/T_e} + r_{\nu,m}^- e^{-\omega_\nu /T_m} \right)}}
\end{align}
where we defined dimensionless relative rates:
\begin{align}
    r_{\nu,j}^-= \frac{\gamma^-_{\nu,j}}{\gamma^-_{\nu,m}+\gamma^-_{\nu,e}},\qquad j=e/m.
\end{align}

We remark that $T^*_\nu$ in general depends on the mode $\nu$ thus giving an overall non-thermal steady state for the multimode cavity.

The result in Eq. \eqref{eq:Teff} can also be found by calculating the Wigner function of the photon state and solving for the steady state, as detailed in App. \ref{app:limblad_solution}. 

\begin{figure}[t]
\centering
    \includegraphics[width=0.95\linewidth]{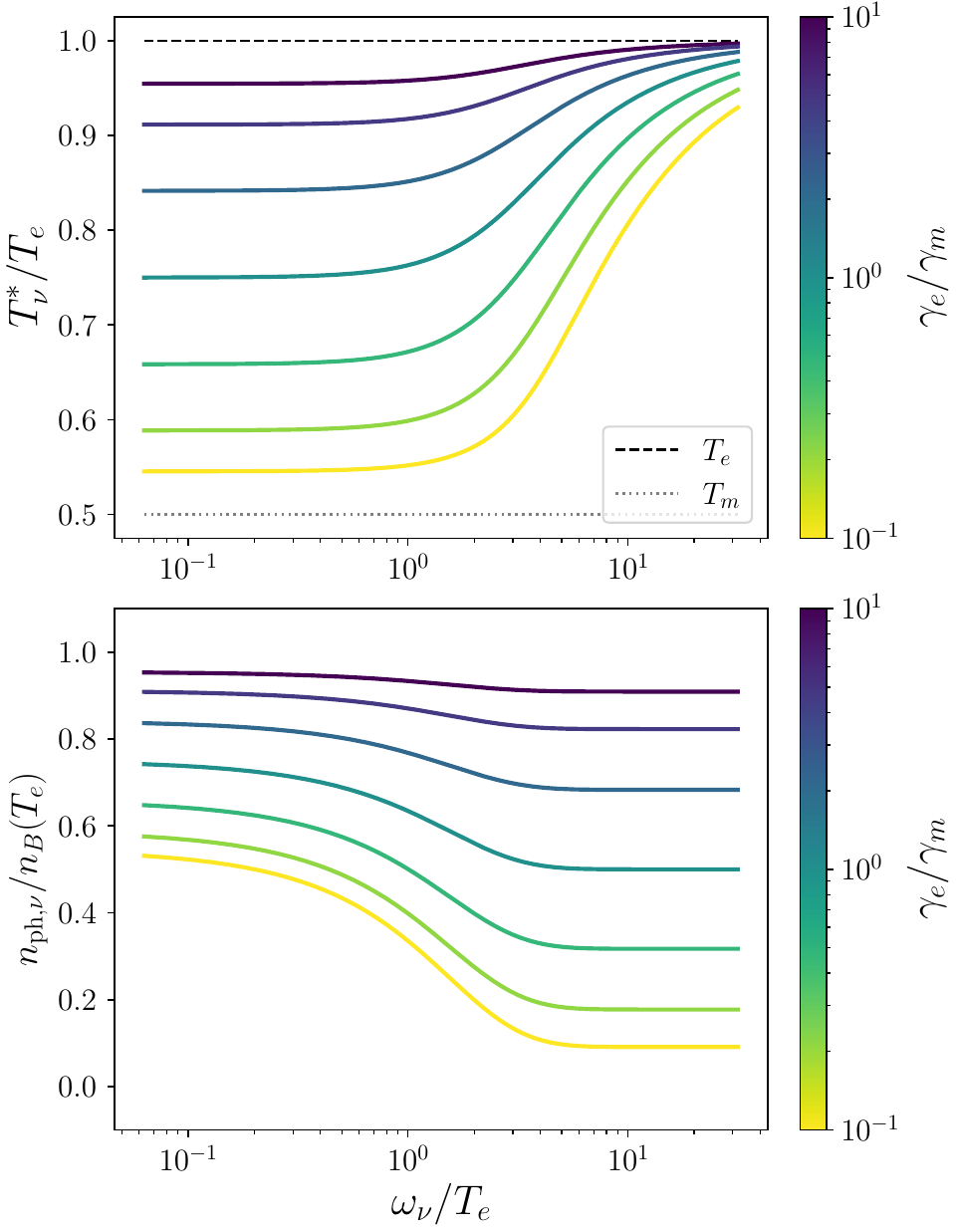}
    \setlength{\unitlength}{1cm}
    \begin{picture}(0,0)
    \put(-7.95,9.5){a)}
    \put(-7.95,4,8){b)}
   \end{picture} 
    \caption{Mode dependent effective temperature $T^*_\nu$ (a) and occupation $n_{\mathrm{ph},\nu}$ (b) for different constant bath coupling rates $\gamma_{\nu,e/m}=\gamma_{e/m}$ as a function of the mode frequency $\omega_\nu$. In particular lines go from electromagnetic enviroment dominated $\gamma_{e}/\gamma_{m}=10$ (blue) to mirror dominated $\gamma_{\nu,e}/\gamma_{\nu,m}=0.1$ (yellow). The mirror temperature is taken as half the electromagnetic enviroment temperature $T_m=0.5 T_e$. Note that the bosonic occupation is shown relative to the occupation at a fixed temperature $T_e$ which becomes exponentially small when $\omega_\nu/T_e\gg 1$.  }
    \label{fig:Teff_nav}
\end{figure}
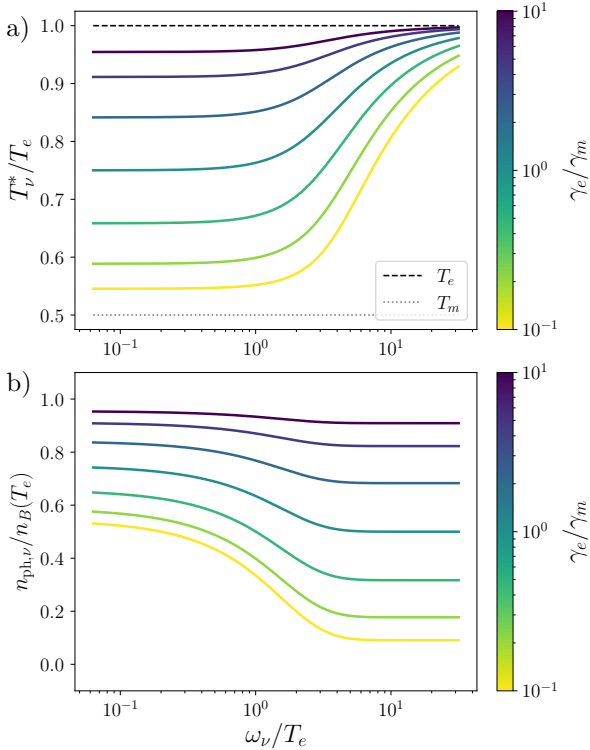

The mode properties can also be characterized by the average occupation number $n_{\mathrm{ph},\nu}=\langle a_\nu^\dagger a_\nu\rangle$ and the probability $p_{n,\nu}$ to have $n$ photons populating mode $\nu$, defined as
\begin{equation}\label{Eq:occupation}
n_{\mathrm{ph},\nu}=\langle a_\nu^\dagger a_\nu\rangle=\text{Tr}(a_\nu^\dagger a_\nu\rho_\nu);\quad p_{n,\nu}=(\rho_\nu)_{nn}.
\end{equation}

From Eqs. \eqref{Eq:limblad_1mode}-\eqref{eq:Teff} it follows that each mode $\nu$ displays thermal properties with temperature $T_\nu^*$, so that $\langle a_\nu^\dagger a_\nu\rangle=n_B(T_\nu^*)$ and $p_{n,\nu}=e^{-n\omega_\nu/T_\nu^*}/(1-e^{-\omega_\nu/T^*_\nu})$. 

We also comment on the relaxation of the cavity to its steady state. For simplicity, one can look at the relaxation rate of a coherence $\langle\hat{a}_\nu \rangle_{t=0}=\alpha_0$ for each mode $\nu$. As we find in App. \ref{app:limblad_solution}, this relaxes with a combined rate $\gamma_{\nu,e}+\gamma_{\nu,m}$, independently of the effective temperature reached by the mode $\nu$ or the temperature of the individual baths. Moreover, for typical cavities, the total dissipation rate $\gamma_{\nu,e}+\gamma_{\nu,m}$ is in the range of MHz-GHz, meaning that relaxation to the steady state occurs on very fast timescales. This justifies studying directly the steady state properties and neglecting the transient dynamics.

We now more explicitly discuss some interesting scenarios for the mode dependent effective temperature $T^*$.

\paragraph{Fabry-Perot cavity. --} The behavior of the effective temperature $T^*_\nu$ behavior as a function of the mode frequency $\omega_\nu$ is plotted in Fig. \ref{fig:Teff_nav} for different values of the ratio $\gamma_e/\gamma_m$, which we assume to be constant in frequency. As we have seen in Section \ref{Sec:RatesExample}, this is a good description of a Fabry-Perot cavity. The effective temperature exhibits two important regimes, namely a \textit{classical} limit ($\omega_\nu \ll T_e,T_m$) and a \textit{quantum} limit ($\omega_\nu\gg T_e,T_m$). In these cases, assuming $T_e>T_m$ and expressing $T^*_\nu$ in terms of the temperature independent rates $\gamma_{\nu,e/m}$, we have:
\begin{flalign}
\label{Eq:TlimCl}    &\textit{classical :}\;\; \;T_{\nu}^{*} \simeq \frac{\gamma_{\nu,e}T_e +\gamma_{\nu,m}T_m }{\gamma_{\nu,e}+\gamma_{\nu,m}};\\
\label{Eq:TlimQu}    &\textit{quantum :}\;\; T_{\nu}^{*} \simeq T_e -\frac{T_e^2}{\omega_\nu}\log\left(\frac{\gamma_{\nu,e}+\gamma_{\nu,m}}{\gamma_{\nu,e}}\right).
\end{flalign}
The classical limit simply recovers a weighted average between the two thermal baths, while in the quantum regime the effective temperature is dominated by the largest temperature, which we assume to be $T_e$. The two regimes are connected by a crossover region, in which $T_\nu^*$ varies smoothly from the classical to the quantum value; this region occurs for frequencies of the order of $T_e$.

In Fig. \ref{fig:Teff_nav}b), we also plot the average photon occupation $\langle \hat{a}^\dagger_\nu \hat{a}_\nu\rangle$ of each mode $\nu$, compared to the thermal occupation at the environment temperature $n_B(T_e)$. The behavior in the two limiting regimes is:
\begin{flalign}
        &\textit{classical :}\;\; \;\langle \hat{a}^\dagger_\nu \hat{a}_{\nu}\rangle\simeq \frac{T^*_\nu}{T_e}n_B(T_e);\\
   &\textit{quantum :}\;\; \langle \hat{a}^\dagger_\nu \hat{a}_{\nu}\rangle \simeq \frac{\gamma_{\nu,e}}{\gamma_{\nu,e}+\gamma_{\nu,m}}n_B(T_e).
\end{flalign}
%
In the classical limit we simply recover a thermal classical behavior, i.e. occupation proportional to $T_\nu^*$. In the quantum limit, $\langle \hat{a}^\dagger_\nu \hat{a}_\nu\rangle$ is always less than that of a thermal state at $T_e$ and also smaller than the occupation in the classical limit, despite $T_\nu^*$ being approximately $T_e$. This is due to the exponential suppression of the occupation in the quantum limit, which amplifies small temperature differences. For intermediate frequency modes, $\langle \hat{a}^\dagger_\nu \hat{a}_\nu\rangle$ is always smaller than the thermal value at $T_e$, due to the logarithmic correction to $T_\nu^*$ in Eq. \eqref{Eq:TlimQu}.

\begin{figure}[t]
    \centering
    \includegraphics[width=0.96\linewidth]{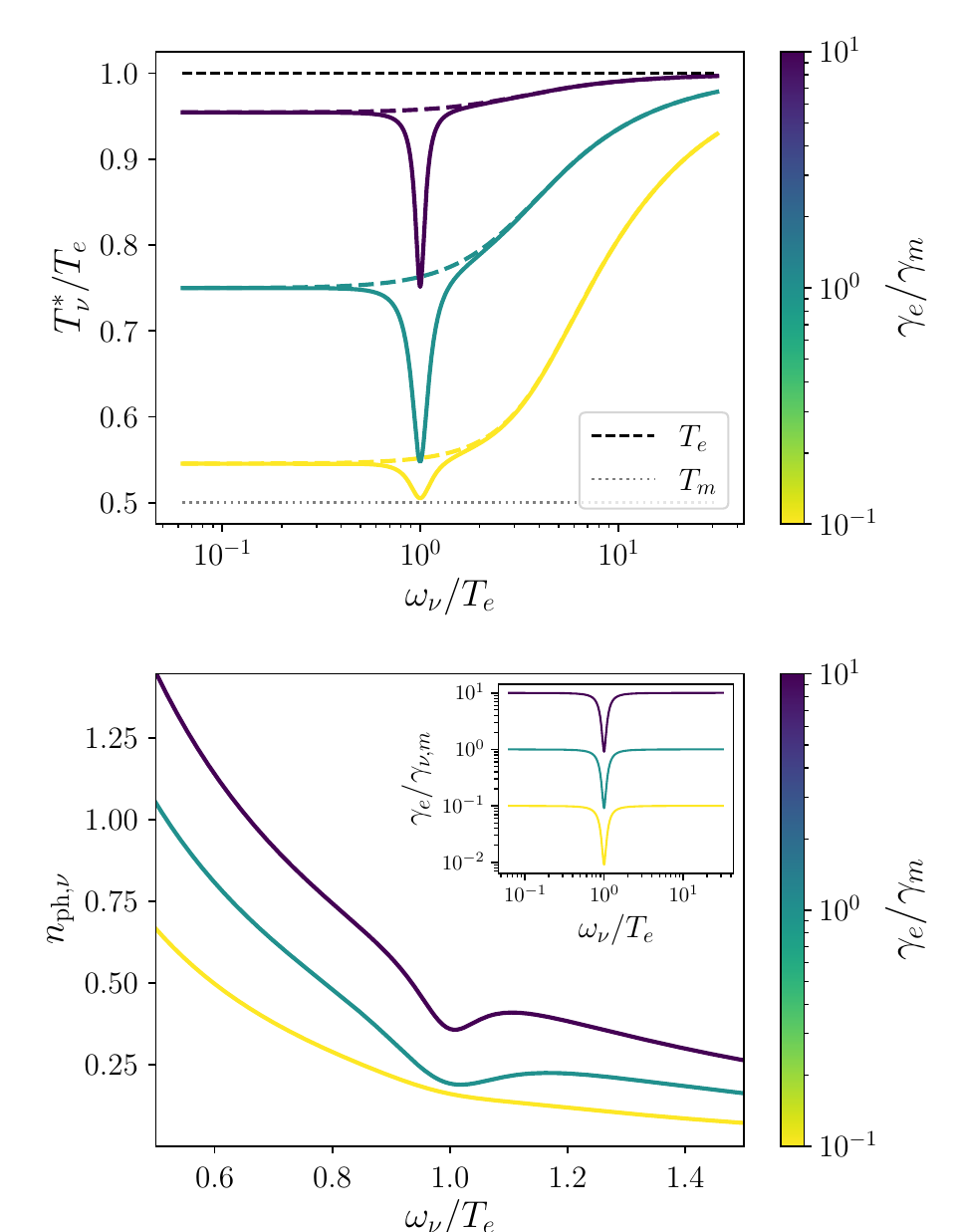}
        \setlength{\unitlength}{1cm}
    \begin{picture}(0,0)
    \put(-7.8,9.3){a)}
    \put(-7.8,4.3){b)}
   \end{picture} 
    \caption{(a) Cavity-mode dependent effective temperature  for frequency-structured mirror dissipation rates as function of the frequency. The mirror dissipation is given by Eq. \eqref{Eq:gamma_mir_RES} with a resonance at $\omega_0/T_e=1$, $1/\tau=0.1T_e$ and $f_{\rm{res}}=10$. The dashed lines indicate the behavior of $T_\nu^*$ if no resonance is present in $\gamma_{\nu,m}$ (i.e. if $f_{\textrm{res}}=0$). (b) Photon occupation number as function of the frequency $\omega_\nu$. A population inversion can be observed near the resonance frequency $\omega_\nu/T_e=1$. The inset shows the behavior of $\gamma_{e}/\gamma_{\nu,m}$.}
    \label{fig:Res}
\end{figure}
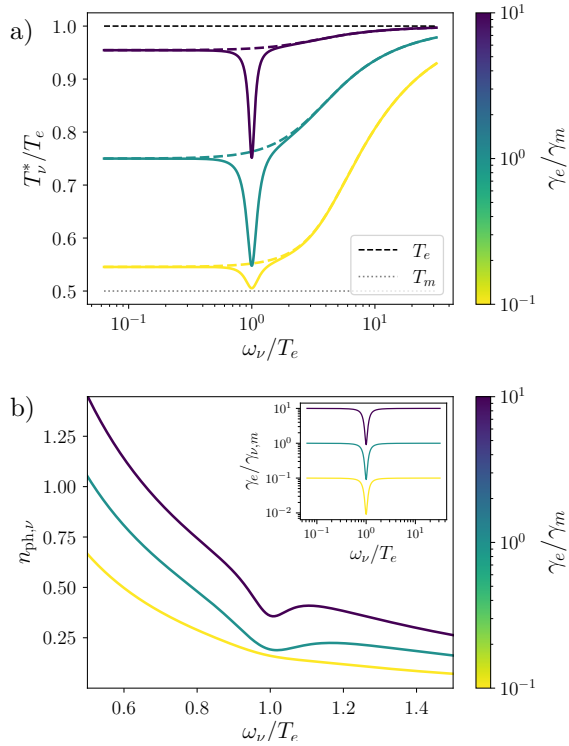

\paragraph{Structured dissipation rates. --} It is in principle possible to shape the frequency dependence of $T_\nu^*$ by engineering structured dissipation rates beyond the simple flat ones ($\gamma_m/\gamma_e=const$) explored above. As an example, we consider the case of a mirror dissipation rate strongly peaked near a specific frequency and show that this simple modification already generates interesting effects. This can occur when the mirrors (or an additional layer) have a strong resonance, leading to a large absorption and a large conductivity in Eq. \eqref{eq:rates_microscopic_mir0}. If the resonance is peaked and narrow enough, dissipation is dominated by the mirrors near the resonance; thus, the temperature of the photons is very close to that of the mirrors near the resonance, and behaves as in Fig. \ref{fig:Teff_nav} away from the resonance.

For the sake of concreteness, we assume that the dissipation rate of the mirrors has a frequency independent term, plus an absorption term, which we model using the Lorentz oscillator model :
\begin{equation}\label{Eq:gamma_mir_RES}
\gamma_{\nu,m}=\gamma_{m}\left(1+f_{\rm{res}}\frac{\omega_\nu^2/\tau^2}{(\omega_\nu^2-\omega_0^2)^2+\omega_\nu^2/\tau^2}\right),
\end{equation}
where we employ $\omega_0/T_e=1$ for the resonance frequency, $1/\tau=0.1\omega_0$ for the width of the resonance and $f_{\rm{res}}=10$ for strength of the resonance. We also assume a frequency independent $\gamma_e$ and study the behavior of the effective temperature for different values of $\gamma_e/\gamma_m$ and for $T_m=0.5T_e$.

In general, the Lindblad equation is not valid when the spectral density of the reservoir (and thus the rates) is frequency dependent \cite{THORWART2004:structured,Weiss2008:structured,Tamascelli2018:structured}. Nonetheless, it is a good approximation when the dissipation rate is smaller than the typical scale on which the spectral density changes, i.e. when $\gamma_{\nu,m}\ll1/\tau$, so that the cavity sees a flat spectral density. We assume that we are in this regime, which is a good approximation for many realistic setups.

Since $\gamma_{\nu,m}$ is peaked around $\omega_0$, the ratio $\gamma_{e}/\gamma_{\nu,m}$ drops around $\omega_0$, causing a drop of $T_\nu^*$ around $\omega_0$, see Fig. \ref{fig:Res}a). This narrow dip in temperature leads to a dip in the energy behavior of the photon population and to a local population inversion, see Fig. \ref{fig:Res}b). The effects of the resonant increase in $\gamma_{\nu,m}$ are stronger when $\gamma_e/\gamma_m$ is large, since for small values of $\gamma_e/\gamma_m$ the temperature of the modes is already close to $T_m$ and further modifications are small. In fact, there is no local population inversion in this case, as shown by the yellow curve in Fig. \ref{fig:Res}b).

These results showcase the possibility to engineer the properties of the photons inside the cavity by manipulating the frequency dependence of  mirrors dissipation rate. Such effect can be easily achieved, for example, by adding specific absorptive layers to the mirrors, and allows to design cavities with highly non-thermal photon states. This effect could provide a novel route to better control quantum materials  \cite{chiriacoThermalPurcellEffect2024,fassioli_controlling_2024,flores-calderonNonthermalElectronphotonSteady2025}, atomic condensates \cite{mixaEnhancingExoticQuantum2024}, chemical reactions \cite{brawleyVibrationalWeakStrong2025} or other systems coupled to the cavity. In particular, highly non-thermal cavity photons can have a dramatic impact on systems very sensitive to radiative energy transfer -- e.g. non-equilibrium metal-insulator transitions \cite{chiriaco_voltage-induced_2018,HanResSwitching2018,Matthies2018Neq} -- or systems where some internal energy relaxation processes are forbidden by symmetry -- e.g. topological materials interacting with polarized light \cite{mciverLightinducedAnomalousHall2020,okaPhotovoltaicHallEffect2009,DagRokajTopologyChiral2024}.

\section{Non-linear mode mixing}\label{Sec:nonLin}

In the previous sections, we have considered a purely quadratic Hamiltonian for the photons. However, realistic cavities may also exhibit coupling between different photon modes, originating from a nonlinear response of the cavity. These nonlinear interactions can arise from various mechanisms, including intrinsic material non-linearities (such as second-order susceptibility $\chi^{(2)}$ in dielectric cavities), effective non-linearities in plasmonic nanocavities, or parametric coupling induced by external driving fields. The presence of such nonlinearities may lead to modifications of  the photon statistics and energy distribution, since now the modes are not independent of each other and one could expect a thermalization process between them.

To account for nonlinear effects in the cavity, we consider a cubic interaction term that couples two photon modes at a time. Specifically, we consider an interaction where two photons from a mode $\nu$ with frequency $\omega_\nu$ can be destroyed to create a photon in mode $2\nu$ with frequency $\omega_{2\nu}=2\omega_\nu$, and vice versa. This process is characteristic of second harmonic generation or other non-linear mixing effects in electromagnetic systems.

The total Hamiltonian now consists of the free Hamiltonian of the cavity modes and the nonlinear interaction term $H = \sum_{\nu} \omega_{\nu} a_{\nu}^\dagger a_{\nu} + H_{\text{int}}$ with
\begin{equation}\label{Eq:HNL_int}
H_{\text{int}} = \sum_{\nu} g_{\nu} \left( a_{2\nu}^\dagger a_{\nu} a_{\nu} +  a_{\nu}^\dagger a_{\nu}^\dagger a_{2\nu}\right),
\end{equation}

 where $g_{\nu}$ is the non-linear coupling coefficient, which depends on the specific physical mechanism generating the interaction. The first term in $H_{\text{int}}$ describes the annihilation of two photons in mode $\nu$ and the simultaneous creation of a photon in mode $2\nu$; the second term is its Hermitian conjugate, representing the reverse process in which a photon in mode $2\nu$ decays into two photons in mode $\nu$.

In order to investigate the effects of the nonlinear interaction on the photon distribution and on the thermal properties of the cavity modes, we restrict our analysis to three specific modes: \(\nu_1 = \nu\), \(\nu_2 = 2\nu\), and \(\nu_3 = 4\nu\). In particular we choose $\nu$ such that $\omega_{\nu_1}/T_e=1$ and $\gamma_m^-=\gamma_e^-$. Without any non-linear interaction, the temperatures of the three modes are $T_1(g=0)/T_e=0.76$, $T_2(g=0)/T_e=0.79$ and $T_3(g=0)/T_e=0.86$.
The Hamiltonian of this reduced system, including the free evolution and the nonlinear coupling terms, is given by  
\begin{equation}\label{Eq:3modes}
H = \sum_{i} \omega_{\nu_i} a_{\nu_i}^\dagger a_{\nu_i} +  g \left( a_{\nu_2}^\dagger a_{\nu_1}^2 + a_{\nu_3}^\dagger a_{\nu_2}^2 + \text{h.c.} \right),
\end{equation}
where \( g \) is assumed to be constant for simplicity. 

To fully capture the dynamics of the interacting modes, we solve for the steady-state density matrix \(\rho\). The presence of nonlinear interactions introduces entanglement between the modes, making a factorization of $\rho$ impossible. Thus, we numerically solve for the full density matrix \(\rho\) by integrating the Lindblad master equation using the QuTiP package \cite{johanssonQuTiPOpensourcePython2012,johanssonQuTiP2Python2013}  
\begin{equation}\label{Eq:3modesLind}
\frac{d\rho}{dt} = -\frac{i}{\hbar} [H, \rho] + \sum_{\substack{i=1,2,3 \\ c=e,m} } \mathcal{L}_{\nu_i,c} (\rho),
\end{equation}
where \( \mathcal{L}_{\nu_i,c}(\rho) \) is the dissipative term associated to photon mode \( \nu_i \) and thermal reservoirs $c$.  

We truncate the Hilbert space by imposing a maximum photon number \( m_i \) for each mode \( \nu_i \). The total Hilbert space is thus of dimension $(m_1+1)(m_2+1)(m_3+1)$, where the truncation parameters \( m_1, m_2, m_3 \) are chosen sufficiently large that the probability to have more than $m_i$ photons in mode $\nu_i$ is negligible. Specifically, we choose $m_1=14$, $m_2=6$ and $m_3=3$. This truncation allows us to perform exact diagonalization of the dynamics of the system while keeping the numerical simulations feasibile.  

From the steady-state solution of \( \rho \), we extract the average photon occupation number of each mode $\langle n_{\nu_i} \rangle $ and the probability that mode $\nu_i$ is occupied by $n$ photons, which are defined by generalizing Eq. \eqref{Eq:occupation}
\begin{gather}
\langle n_{\nu_i} \rangle = \text{Tr} \left( a_{\nu_i}^\dagger a_{\nu_i} \rho \right);\quad p_{n,\nu_i}=[\text{Tr}_{j\neq i}(\rho)]_{nn}.
\end{gather}

\begin{figure}[t]
    \centering
    \includegraphics[width=\linewidth]{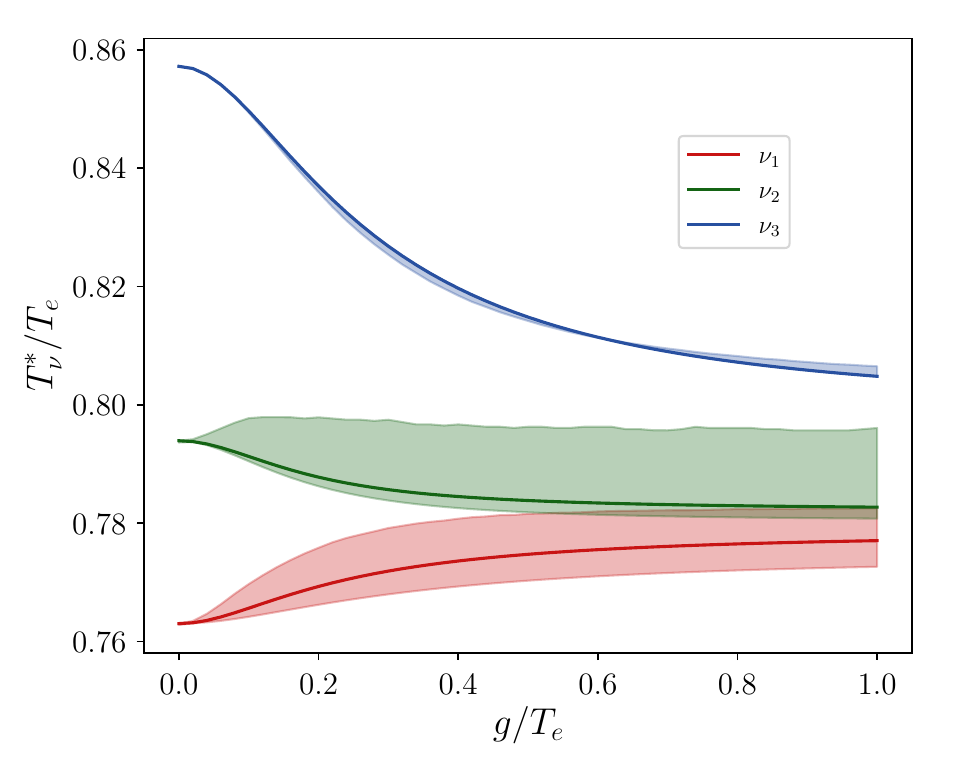}
    \caption{Effective temperature $T_\nu^*$ for each of the three modes considered, as function of the non linear coupling g. The shaded regions indicate the interval of the effective temperature, due to deviations from thermal statistic for each mode.}
    \label{fig:3modes}
\end{figure}

We find that the population of each mode is not thermal, in contrast to the quadratic case of  Sec. \ref{Sec:Solution} where $p_{n,\nu}\sim e^{-n\omega_\nu/T_\nu^*}$. In order to study the properties of the photons we perform two analysis. \emph{(i)} Calculate an average effective temperature $T_{\nu_i}$ by fitting the population of each mode with a Bose-Einstein distribution, $\langle n_{\nu_i} \rangle = n_B(\omega_{\nu_i}/T_{\nu_i})$. \emph{(ii)} Perform a linear fit of $\ln p_{n,\nu_i}$ as function of $n$, in order to estimate how much $p_{n,\nu_i}$ deviates from a thermal statistics and calculate the maximum and minimum effective temperatures that give the temperature range.

The results are plotted in Fig. \ref{fig:3modes} as a function of \( g \), showing the effect of nonlinear interactions on the average effective temperature of each mode, and on the maximum and minimum temperature. We observe that for small \( g \), the effective temperatures remain close to their $g=0$ values dictated by the external reservoirs, and the range given by the minimum and maximum temperatures is small, so that the statistic of each mode remains quasi-thermal. However, as \( g \) increases, the nonlinear mode mixing terms increase thermalization between the different modes and their temperatures get closer to each other. The deviation of $p_{n,\nu}$ from a thermal statistic is larger for stronger $g$ and for the two lower modes, an effect qualitatively similar to the Raman parametric amplification found in \cite{ojedacolladoEquilibriumParametricAmplification2024}. Nonetheless, the non-linear terms have an overall weak effect, which is negligible in realistic situations, where $g$ is small compared to the typical energy scales of the cavity.

\section{Conclusions}\label{Sec:Conclusions}

In this work, we investigated the dissipative dynamics and the (non) thermal properties of photons in a cryogenic electromagnetic cavity. We formulated a general description based on a Lindblad master equation, accounting for dissipation arising from the coupling to two thermal reservoirs at distinct temperatures (the cryogenically cooled mirrors and the warmer external environment). Our approach provides a clear method to calculate the properties of the cavity modes and estimate the relevance of thermal effects.

Our analysis reveals that each cavity mode displays a steady state characterized by a thermal statistics with a frequency-dependent effective temperature. The effective temperature exhibits a crossover from a low frequency classical regime, where it is a weighted average of the two reservoir regimes, to a high frequency quantum regime, where the cavity thermalizes to the warmer reservoir temperature.

We also showed that the frequency dependence of the dissipation rates can be exploited to tailor the properties of the cavity photons, for example modifying their energy distribution and creating a local population inversion. These results offer a new avenue for cavity engineering and for the control of materials embedded in the cavity. For example, it has been shown \cite{flores-calderonNonthermalElectronphotonSteady2025} that the electronic properties of a material can already be manipulated in a non invasive way by coupling it to a cryostat and to a cavity at different temperatures. Introducing a frequency structured photonic environment adds a new control parameter available for manipulating the material properties.

Our results highlight that extracting the total dissipation rate from measurements of the quality factor alone is insufficient, as the individual dissipation rates for mirrors and environment are needed when they have different temperatures. Thus, we derived their explicit expressions from a microscopic model, and linked them to measurable macroscopic quantities such as conductivity and geometry of the mirrors. The expressions we derived hold for a generic cavity formed by a confining matter medium and by an external photonic environment, and only require the knowledge of a few experimental parameters. We performed explicit calculations for the case of a Fabry-Perot cavity and of a nanoplasmonic cavity. 

Furthermore, we examined the effects of a nonlinear coupling between cavity modes. These interactions only cause weak modifications to the effective temperature, without inducing full thermalization, thus showing that our results are also valid for cavities with inter-mode interactions.

Our findings have important implications for experiments involving cryogenic cavities, particularly those investigating cavity-modified material properties. In particular, we provide simple expressions to calculate the dissipation rates from macroscopic properties of the cavity, which are all experimentally accessible, allowing researchers to determine when dissipation and thermal effects are significant. Our results also provide a theoretical foundation to optimize experimental cavity setups, mitigate thermal effects, or even engineer the properties of the cavity photons.

\paragraph{Acknowledgments. ---} 

We thank G. M. Andolina, M. Dalmonte, D. De Bernardis and F. Pellegrino for fruitful discussions. G.P., E.P. and G. F. acknowledge the PNRR MUR Project PE0000023-NQSTI. G.C. is supported by ICSC – Centro Nazionale di Ricerca in High-Performance Computing, Big Data and Quantum Computing under project E63C22001000006.

\bibliography{biblio.bib}

\onecolumn\newpage
\appendix

\section{Electric field in the energy eigenbasis}
\label{app:eigenoperators}
Let us provide a derivation of the expression of the electric field in the energy eigenbasis. A generic operator $A$ can be written in the energy eigenbasis as $A(\omega) = \sum_{\varepsilon- \varepsilon' = \omega}\Pi(\varepsilon)A\Pi(\varepsilon')$, being $\Pi(\varepsilon)$ the projector in the eigenspace belonging to the eigenvalue $\varepsilon$. 
In our case, the system operators are the field operators $a^{(\dagger)}_\nu$. As orthonormal basis we choose the Fock one $\{ \ket{\mbf n} =\ket{n_1}\otimes\ket{n_2}\otimes \dots \otimes\ket{n_N}\}$, with $n_\nu$ the occupation number associated to the mode $\nu$.
We obtain
\begin{equation}
	{\mbf E}_\omega(\mbf r) = \sum_{nm} \delta(\varepsilon_m - \varepsilon_n - \omega) {\mbf E}_{nm}({\mbf r}) \ket{\bf n}\bra{\bf m},
\end{equation}
being ${\mbf E}_{nm}({\mbf r}) = \braket{\bf n| {\mbf E}({\mbf r}) |\bf m}$ and $\varepsilon_n = \sum_\nu^N n_\nu\, \omega_\nu$. 
We can substitute Eq.~\eqref{eq:efield} and find
\begin{equation}
	{\mbf E}_{nm}({\mbf r}) = i\sum_\nu [\kappa_\nu(\mbf r) \braket{\mbf n|a_\nu|\mbf m}-\kappa_\nu^*(\mbf r) \braket{\mbf n|a_\nu^\dagger|\mbf m}],
\end{equation}
with $\kappa_{\nu}(\mbf r) = \sqrt{\frac{\omega_{\nu}}{2\epsilon_0V_{\nu}}}\,{\mbf e}_{\nu}(\mbf r)$. To simplify the notation, we introduce the two terms
\begin{equation} 
{\mbf E}^\text{em}_{nm}({\mbf r}) = i\sum_\nu \kappa_\nu(\mbf r) \braket{\mbf n|a_\nu|\mbf m}\ \ \text{  and  } \ \ {\mbf E}_{nm}^\text{ab}({\mbf r}) = -i\sum_\nu \kappa_\nu^*(\mbf r) \braket{\mbf n|a_\nu^\dagger|\mbf m}.
\end{equation}

Accordingly with this notation, the electric field decomposes as 
\begin{equation}
{\mbf E}_\omega({\mbf r}) = {\mbf E}^\text{ab}_\omega({\mbf r})+{\mbf E}^\text{em}_\omega({\mbf r}),
\end{equation} 
with 
\begin{equation} 
{\mbf E}^\text{ab/em}_\omega({\mbf r})=\sum_{nm} \delta(\varepsilon_m - \varepsilon_n - \omega) {\mbf E}^\text{em/ab}_{nm}({\mbf r}) \ket{\bf n}\bra{\bf m}. 
\end{equation}
We notice that 
\begin{equation}
	\braket{\mbf n|a_\nu|\mbf m} = \sqrt{n_\nu+1}\ \delta_{m_1, n_1} \delta_{m_2, n_2} \dots \delta_{m_\nu, n_\nu+1} \dots \delta_{m_N, n_N},
    \end{equation}
implying $\varepsilon_m - \varepsilon_n = \omega_\nu$. 
Analogously we have 
\begin{equation}
	\braket{\mbf n|a_\nu^\dagger|\mbf m} = \sqrt{n_\nu}\ \delta_{m_1, n_1} \delta_{m_2, n_2} \dots \delta_{m_\nu, n_\nu-1} \dots \delta_{m_N, n_N},
    \end{equation}
implying $\varepsilon_m - \varepsilon_n = -\omega_\nu$.
We can substitute and exploit the identities $a_\nu = \sqrt{n_\nu+1}\ket{n_\nu}\bra{n_\nu+1}$ and $a_\nu^\dagger = \sqrt{n_\nu}\ket{n_\nu}\bra{n_\nu-1}$ and find Eqs.~\eqref{Eq:em_ab}

\section{Relation between conductivity and dissipation rates}
\label{app:diss_conductivity}

In this Appendix we make use of the Lehmann representation to show that
\begin{equation}
	\Gamma(-{\mbf r},- \omega) =e^{-\beta \omega} \Gamma ({\mbf r},\omega).
\end{equation}
This follows directly from the definition, by assuming the mirror in the thermal state
\begin{equation}
    \rho_\text{m} = Z_\text{m}^{-1}\sum_{i} e^{-\beta_\text{m} \epsilon_{i}} \ket{i}\bra{i}.
\end{equation} 
Here, $\beta_\text{m} = T_\text{m}^{-1}$ is the inverse mirror temperature, $\epsilon_{i}$ the energy eigenvalue associated to the $i$-th eigenstate of the mirror Hamiltonian $H_m$ (which we do not need to specify), and $Z_\text{m}=\sum_{i} e^{-\beta_m \epsilon_{i}}$ the partition function. For simplicity of notation we pose $p_{i}=Z_\text{m}^{-1}e^{-\beta_\text{m} \epsilon_{i}}$.
Under this assumption the two-times dipole correlation function reads
\begin{equation}\label{Eq:Lehmann_dd}
	\begin{aligned}
		\braket{{\mbf d}({\mbf r}, \tau) {\mbf d}(0, 0)}  =& \text{Tr}(\rho_\text{m}e^{iH_m\tau}\mbf d(\mbf r)e^{-iH_m\tau}\mbf d)= \sum_{ih} p_{i} e^{-i (\epsilon_h -\epsilon_{i})\tau} d_{ih}({\mbf r})d_{hi}(0);\\
        \braket{{\mbf d}(0,0) {\mbf d}({\mbf r}, \tau)}  = & \sum_{ih} p_{i} e^{-i (\epsilon_i -\epsilon_{h})\tau} d_{ih}(0)d_{hi}({\mbf r})=\sum_{ih} p_{h} e^{-i (\epsilon_h -\epsilon_{i})\tau} d_{ih}(\mbf r)d_{hi}(0).
	\end{aligned}
\end{equation}
where we used the Lehmann representation to write the dipole operator matrix elements.

Accordingly with this expression we can write
\begin{equation}
	\begin{aligned}
		\Gamma({\mbf r},\omega)
        &=\pi\sum_{ih} p_{i} d_{ih}({\mbf r})d_{hi}(0) \delta(\epsilon_h - \epsilon_{i} - \omega). \\
	\end{aligned}
\end{equation}
where we retained the real part of the integral over $\tau$ and used $\int_0^{\infty}d\tau e^{i\omega\tau}=\pi\delta(\omega)+i\mathcal{P}\frac1{\omega}$.

By substituting $\omega$ with $- \omega$ we can easily find
\begin{equation}
\begin{aligned}
	\Gamma(-{\mbf r},- \omega) 
 &=\int_0^{\infty}d\tau e^{i(-\omega)\tau}\braket{{\mbf d}(0, \tau) {\mbf d}({\mbf r}, 0)}=\pi\sum_{ih} p_{i} d_{ih}(0)d_{hi}({\mbf r}) \delta(\epsilon_{i} - \epsilon_h - \omega)=\\
 &=\pi\sum_{ih} p_{h} d_{ih}({\mbf r})d_{hi}(0) \delta(\epsilon_{h} - \epsilon_i - \omega)=e^{-\beta \omega} \Gamma ({\mbf r},\omega).
 \end{aligned}
\end{equation}
The last line follows from the relation $p_h = e^{-\beta \epsilon_h} = e^{-\beta (\epsilon_{i} + \omega) } = p_{i} e^{-\beta \omega}$, after having exchanged $i$ and $h$. 

\begin{gather}
\notag\sigma(\mbf q,\,\omega) = \frac{1}{i} \int_0^{\infty} dt \int d^3\mbf re^{i\omega t-i\mbf q\cdot\mbf r}\braket{[{\mbf d}(\mbf r,t), {\mbf J}]}=\frac{1}{i} \int_0^{\infty} dt \int d^3\mbf re^{i\omega t-i\mbf q\cdot\mbf r}\braket{[{\mbf d}, {\mbf J}(-\mbf r,-t)]}=\\
\notag    =\frac{1}{i} \int_0^{\infty} dt \int d^3\mbf re^{i\omega t-i\mbf q\cdot\mbf r}\braket{[{\mbf d}, -\partial_t{\mbf d}(-\mbf r,-t)]}=\omega \int_0^{\infty} dt \int d^3\mbf re^{i\omega t-i\mbf q\cdot\mbf r}\braket{[{\mbf d},{\mbf d}(-\mbf r,-t)]}=\\
\notag    =\omega\int_0^{\infty} dt \int d^3\mbf re^{i\omega t-i\mbf q\cdot\mbf r}(\braket{{\mbf d}(\mbf r,t){\mbf d}(0,0)}-\braket{{\mbf d}(0,0) {\mbf d}(\mbf r,t)})=\\
\label{Eq:fluct_dis}    =\omega\int d^3\mbf re^{i\omega t-i\mbf q\cdot\mbf r}(\Gamma(\mbf r,\omega)-e^{-\beta_{\rm{m}}\omega}\Gamma(\mbf r,\omega))=\omega\tilde\Gamma(\mbf q,\omega)(1-e^{-\beta_{\rm{m}}\omega})
\end{gather}
where we used Eq, \eqref{Eq:Lehmann_dd} to write the expectation values in terms of $\Gamma$.

From Eq. \eqref{Eq:fluct_dis} we thus show that
\begin{equation}
\tilde\Gamma(0, \omega_{\nu})=\frac{\sigma(\omega_{\nu})}{\omega_{\nu}(1-e^{-\beta_{\rm{m}}\omega_{\nu}})};\qquad \tilde\Gamma(0, -\omega_{\nu})=\frac{\sigma(\omega_{\nu})}{\omega_{\nu}(e^{\beta_{\rm{m}}\omega_{\nu}}-1)}. 
\end{equation}

\section{Calculation of electromagnetic field modes}\label{App:Fields}

In this Appendix we detail out the calculation of the electromagnetic field for the resonant modes of a Fabry-Perot planar cavity and for a metallic surface supporting plasmonic modes.

\subsection{Cavity modes}\label{App:FPcavity}

We consider an ideal Fabry-Perot cavity with infinite planar mirror extending in the $x$
-$y$ plane and centered at $z=\pm L_c/2$. The mirrors have thickness $d$ -- so that they extend within $-(L_c+d)/2<z<-(L_c-d)/2$ and $(L_c-d)/2<z<(L_c+d)/2$ -- and index of refraction $n(\omega)=\sqrt{\varepsilon(\omega)}$. The dielectric constant is approximately $\varepsilon=1-\omega_p^2/\omega^2\approx-\omega_p^2/\omega^2$ with $\omega_p$ the plasma frequency of the metal of the mirrors. In the regime $\omega\ll\omega_p$, the index of refraction is $n(\omega)\approx i\omega_p/\omega$ and electromagnetic waves decay inside the mirror with a characteristic length $\sim c/\omega_p$. If $\omega_pd/c\gg1$ -- as it typically occurs for the mirrors in good quality cavities --

In order to calculate the cavity and free space modes $\mbf e_\nu$ and $\mbf f_{\mbf k,\lambda}$, the correct and formal approach is to solve for the Maxwell equations in the whole space, and then quantize the field inside the cavity and in the external free space \cite{demartiniSpontaneousEmissionOptical1991,viviescasFieldQuantizationOpen2003}. This procedure is rather cumbersome, but can be simplified if the mirrors of the cavity are highly reflective. In this regime, the modes of the cavity field and of the free space field essentially see the mirrors as semi-infinite since $\omega_pd/c\gg1$ and can be treated independently. Therefore, we only need two consider two separate scattering processes: one for the free space modes at the $z=-(L_c+d)/2$ and one for the cavity modes $z=-(L_c-d)/2$ (the case of the mirror at $z=L_c/2$ is identical).

We consider a quantization volume for the free space modes with dimensions $L_x$, $L_y$ and $L_z$. Assuming $k_y=0$ for simplicity (since there is rotational invariance in the $x$-$y$ plane), we can write
\begin{gather}
\mbf f_{\mbf k,\lambda}=\mbf u_{\lambda,+}e^{ik_xx+ik_zz'}+r_{\lambda}\mbf u_{\lambda,-}e^{ik_xx-ik_zz'}\text{   for   }z'<0;\\
\mbf f_{\mbf k,\lambda}=t_{\lambda}\mbf u'_{\lambda,+}e^{ik_xx-i\kappa z'}\text{   for   }z'>0.
\end{gather}
where $\lambda=s,p$ is the TE (TM) polarization, $\mbf k=(k_x,0,k_z)$, $\kappa=\sqrt{n^2\omega^2/c^2-k_x^2}\approx\omega_p/c$, and $z'=z+(L_c+d)/2$ is the shifted $z$ coordinate such that the mirror-free space interface is located at $z'=0$. The polarization vectors are $u_{s,\pm}=u_{s,+}=(0,-1,0)$ and $u_{p,\pm}=(\cos\theta,0,\mp\sin\theta)$ and $u'_{p,+}=(\cos\theta',0,-\sin\theta')$, with $\cos\theta=k_zc/\omega$ and $\cos\theta'=i\kappa c/(n\omega)\approx1$, $\sin\theta'=k_xc/(n\omega)=\sin\theta/n\approx-i\sin\theta\omega/\omega_p$. The reflection and transmission coefficients $r$ and $t$ are given by the standard Fresnel coefficients.
\begin{gather}
r_s=-\frac{n\cos\theta'-\cos\theta}{n\cos\theta'+\cos\theta};\qquad t_s=\frac{2\cos\theta}{n\cos\theta'+\cos\theta};\\
r_p=-\frac{n\cos\theta-\cos\theta'}{n\cos\theta+\cos\theta'};\qquad t_p=\frac{2\cos\theta}{n\cos\theta+\cos\theta'}
\end{gather}

Expanding to the leading order in $\omega/\omega_p$, we find $r_s=-(1-2\cos\theta/(n\cos\theta'))$, $r_p=-(1-2\cos\theta'/(n\cos\theta))$, $t_s=2\cos\theta/(n\cos\theta')$, $t_p=2/n$, so that we can write at once both the free space and cavity modes inside the mirror:
\begin{gather}
\mbf f_{\mbf k,s}=t_{s}\begin{pmatrix}0\\-1\\0\end{pmatrix}e^{ik_xx-\kappa(z+(L_c+d)/2)};\qquad \mbf e_{\mbf q,s}=t_{s}\begin{pmatrix}0\\-1\\0\end{pmatrix}e^{iq_xx-\kappa(d-z-(L_c+d)/2)};\\
\mbf f_{\mbf k,p}=t_{p}\begin{pmatrix}\cos\theta'\\0\\-\sin\theta'\end{pmatrix}e^{ik_xx-\kappa(z+(L_c+d)/2)};\qquad \mbf e_{\mbf q,p}=t_{p}\begin{pmatrix}\cos\theta'\\0\\\sin\theta'\end{pmatrix}e^{iq_xx-\kappa(d-z-(L_c+d)/2)}
\end{gather}

Since conservation of momentum in Eq. \eqref{eq:rates_microscopic_env0} constrains $q_x=k_x$, while conservation of energy constrains $k_z=q_z$, we can restrict ourselves to the case $\mbf f_{\mbf k,\lambda}\cdot\mbf e_{\mbf k,\lambda'}$. Using the fact that $\sin\theta'^*\approx-\sin\theta$ and that different polarizations are orthogonal, we find
\begin{equation}
\mbf f_{\mbf k,\lambda}\cdot\mbf e_{\mbf k,\lambda'}=\delta_{\lambda,\lambda'}|t_{\lambda}|^2e^{-\kappa d}
\end{equation}
Therefore in Eq. \eqref{eq:rates_microscopic_env0} we set $\nu=q_x,m,\lambda$, where $q_x$ indicates the transverse momentum, $m$ the quantum number associated to the longitudianal momentum $q_z=m\pi/L_c=m\omega_c/c$ and $\lambda$ the polarization. The modes volumes are given by $V=L_xL_yL_z$ and $V_{\nu}=2L_xL_yL_c$. Using $\omega_\nu=\sqrt{c^2q_x^2+(m\omega_c)^2}$ and accounting for both mirrors, we can write
\begin{gather}
\gamma_{\nu,e}=\sum_{\mathbf{k}, \lambda} \frac{\pi \omega_{\nu}^2}{2V V_{\nu}} \left| \mathcal{A}_{\mbf k,\lambda} \right|^2 \delta(\omega_{\nu} - \omega_\mbf k)\\
\mathcal{A}_{\mbf k,\lambda}=\int d^3\mathbf{r} \,\varepsilon(\mbf r) \mathbf{f}_{\mathbf{k}, \lambda}(\mathbf{r}) \cdot \mathbf{e}_{\nu}^*(\mathbf{r})=\delta_{q_x,k_x}L_xL_y \int_0^{d}dz \,n^2| t_{\lambda}|^2e^{-\kappa d}\\
    \gamma_{\nu,e}=\frac{\pi \omega_{\nu}^2}{2}\sum_{k_x}\frac{(\delta_{q_x,k_x}L_xL_y)^2}{(L_xL_y)^22L_c} \int_0^{\infty}\frac{dk_z}{2\pi}\left| \int_0^{d}dz \,n^2| t_{\lambda}|^2e^{-\kappa d} \right|^2 \delta(\omega_{\nu} - \omega_\mbf k)=\\
    =\frac{\pi \omega_{\nu}^2}{2 (2L_c)} \int_0^{\infty}\frac{dk_z}{2\pi}d^2e^{-2\kappa d}|nt_{\lambda}|^4 \frac1c\delta\left(\sqrt{q_x^2+(m\pi/L_c)^2} - \sqrt{q_x^2+k_z^2}\right)=\\
=\frac{\omega_\nu^2\omega_c}{8\pi}\frac{d^2}{c^2}e^{-2\kappa d}\frac{\omega_\nu}{m\omega_c}|nt_{\lambda}|^4
\end{gather}

So that we find
\begin{equation}\label{Eq:gamma_env_FPC_app}
\gamma_{\nu,s,e}=\frac{\omega_\nu^2\omega_c}{2\pi}\frac{d^2}{c^2}e^{-2\kappa d}\left(\frac{m\omega_c}{\omega_\nu}\right)^3;\qquad \gamma_{\nu,p,e}=\frac{\omega_\nu^2\omega_c}{2\pi}\frac{d^2}{c^2}e^{-2\kappa d}\frac{\omega_\nu}{m\omega_c}
\end{equation}
Since $\omega_\nu=\sqrt{c^2q_x^2+(m\omega_c)^2}$ the last factors are typically of order one except for modes with high transverse momentum. In the limiting case of the TM$_0$ mode ($m=0$, $\lambda=p$), the dissipation rate vanishes because there is no transmitted wave inside the mirrors, i.e. $\gamma_{m=0,p,e}=0$.

The dissipation rates associated to the mirrors can be easily calculated using $\int_mdz|\mbf e_{\nu}(\mbf r)|^2\approx\frac{1}{2\kappa}|t_{\lambda}|^2$, so that
\begin{equation}\label{Eq:gamma_mir_FPC_app}
\gamma_{\nu,s,m}=\frac{\sigma(\omega_\nu)}{\pi\epsilon_0}\frac{\omega_\nu^2}{\omega_p^2}\frac{\omega_c}{\omega_p}\frac{m\omega_c}{\omega_\nu};\qquad \gamma_{\nu,p,m}=\frac{\sigma(\omega_\nu)}{\pi\epsilon_0}\frac{\omega_\nu^2}{\omega_p^2}\frac{\omega_c}{\omega_p}
\end{equation}

\subsection{Plasmonic modes}\label{App:Plamonics}

A metallic substrate occupies the half space at $z<0$, while the space above it is occupied by air or vacuum(with dielectric constant equal to 1). The metal has a complex dielectric constant $\varepsilon(\omega)=\varepsilon'+i\varepsilon''=1-\frac{\omega_p^2}{\omega^2}+i\frac{\sigma(\omega)}{\epsilon_0\omega}$, where $\sigma(\omega)$ is the conductivity and $\omega_p$ is the plasma frequency. We assume $\omega/\omega_p\ll1$ and calculate all quantities of interest to the leading order.

We assume invariance along the $y$ direction. This configurations supports localized plasmonic modes propagating along the $x$ direction and exponentially decaying away from the metallic surface. Accounting for losses in the metal yields decay in the $x$ direction as well. The solution of the Maxwell equations for a plasmon with frequency $\omega$ is
\begin{equation}\label{Eq:enuPlasmApp}
\mbf e_\omega(\mbf r,t)=\begin{pmatrix}i\kappa_1/\omega\\0\\(\mp\kappa_1/\kappa_j)q_x/\omega\end{pmatrix}e^{iq_xx-\kappa_x|x|-\kappa_j|z|-i\omega t}
\end{equation}

Here the $\mp$ sign and $j=1,2$ refer to the $z>0$ ($z<0$) halfspace. The wavevector along $x$ is given by $q_x$, with $\kappa_x\ll q_x$ the decay component of the wavevector, while $\kappa_j$ express the decay of the field in the $z$ direction. Note that the field satisfies $\nabla\cdot\mbf e_\omega=0$ everywhere except at $x=0$, which means that the plasmon is excited by a suitable pump at $x=0$.

Imposing the boundary conditions on the electric field and the dispersion relation, we obtain
\begin{gather}
\notag
\kappa_2=\varepsilon'\kappa_1;\quad (q_x+i\kappa_x)^2-\kappa_1^2=\omega^2;\quad (q_x+i\kappa_x)^2-\kappa_1^2=\varepsilon\omega^2\\
\label{Eq:PlasmonDispersion}
q_x^2=\frac{\varepsilon'}{1+\varepsilon'}\omega^2;\qquad \kappa_x=\frac{\varepsilon''}{(1+\varepsilon')^2}\frac{\omega^2}{2q_x}; \qquad \kappa_1^2=-\frac1{1+\varepsilon'}\omega^2;\qquad \kappa_2^2=-\frac{\varepsilon'^2}{1+\varepsilon'}\omega^2\\
\label{Eq:PlasmonDispersion2}
q_x\approx\omega+\frac{\omega^3}{2\omega_p^2};\qquad \kappa_x\approx\frac{\varepsilon''}{2}\left(\frac{\omega}{\omega_p}\right)^4\omega; \qquad \kappa_1\approx\frac{\omega^2}{\omega_p};\qquad \kappa_2\approx\omega_p
\end{gather}

It is important to retain the correction to $q_x$ because it would otherwise lead to a divergence in the calculation of $\gamma_e$. The presence of a finite $\kappa_x$ is also required to couple to the free space field. In fact, for $\kappa_x=0$ a free space mode with $\mbf k=(k_x,0,k_z)$ needs to conserve the energy and the momentum along $x$ in order to couple to the plasmon, meaning that $\omega^2=q_x^2+k_z^2$, which has no solution since $q_x>\omega$. On the other hand, if $\kappa_x\neq0$ a free space mode with $k_x<\omega<q_x$ can still couple to the plasmonic mode, although with a reduced overlap.

The mode volume is the sum of two contributions: the one from the field inside the metal and the contribution from the field outside of the metal. The first term in particular is
\begin{gather}
\int_md\mbf r|\mbf e_{\omega}(\mbf r)|^2=L_y\int dxe^{-2\kappa_x|x|}\int_0^{\infty}dz e^{-2\kappa_2z}\frac{q_x^2\kappa_1^2/\kappa_2^2+\kappa_1^2}{\omega^2}\approx\\
\approx\frac{L_y}{\kappa_x}\frac1{2\kappa_2}(\omega^4/\omega_p^4+\omega^2/\omega_p^2)\approx\frac{L_y}{2\kappa_x\kappa_1}\left(\frac{\omega}{\omega_p}\right)^4
\end{gather}
\begin{gather}
V_{\omega}=L_y\int dxe^{-2\kappa_x|x|}\int_0^{\infty}dze^{-2\kappa_1z}\frac{q_x^2+\kappa_1^2}{\omega^2}+\varepsilon'\int_md\mbf r|\mbf e_{\omega}(\mbf r)|^2\approx\\
\approx\frac{L_y}{\kappa_x}\left(\frac1{2\kappa_1}(1+\omega^2/\omega_p^2)-\frac1{2\kappa_2}(1+\omega^2/\omega_p^2)\right)\approx\frac{L_y}{2\kappa_x\kappa_1}
\end{gather}
In other words most of the mode volume is given by the field living outside of the metal, as one would expect. 

We can then calculate the dissipation rate $\gamma_m$ as
\begin{equation}
\gamma_{\omega,m}=\frac{\sigma(\omega)}{2\epsilon_0}\left(\frac{\omega}{\omega_p}\right)^4=\frac\omega2\frac{\sigma(\omega)}{\epsilon_0\omega_p}\left(\frac{\omega}{\omega_p}\right)^3
\end{equation}

The free space modes can be calculated by considering the standard problem of a plane wave with $\mbf k=(k_x,0,-k_z)=k(\sin\theta,0,-\cos\theta)$ hitting the metallic surface and generating a reflected and a transmitted wave. Solving to the leading order in $\omega/\omega_p$ we find
\begin{equation}\label{Eq:fkApp}
    \mbf f_\mbf k(\mbf r)=e^{ik_xx}\left(e^{-ik_zz}\begin{pmatrix}
        \cos\theta\\ 0 \\\sin\theta
    \end{pmatrix}+re^{ik_zz}\begin{pmatrix}
        -\cos\theta\\ 0 \\\sin\theta
    \end{pmatrix}\right);\qquad \mbf f_\mbf k(\mbf r)=e^{ik_xx}e^{\kappa z}2\frac{\omega}{\omega_p}\begin{pmatrix}
        i\\ 0 \\\sin\theta\frac{\omega}{\omega_p}
    \end{pmatrix}
\end{equation}

where $r\approx1-2i\frac{\omega}{\omega_p\cos\theta}$ is the reflection coefficient and $\kappa\approx\omega_p$.

We can now calculate the overlap integral $\int d\mbf r\varepsilon\mbf f_\mbf k(\mbf r)\cdot\mbf e_\nu(\mbf r)$. First we obtain 
\begin{gather*}
\mbf f_\mbf k(\mbf r)\cdot\mbf e_\nu(\mbf r)=e^{i(k_x-q_x)x-\kappa_x|x|}[e^{-(\kappa_1+ik_z)z}(\sin\theta-i\omega/\omega_p\cos\theta)+re^{-(\kappa_1-ik_z)z}(\sin\theta+i\omega/\omega_p\cos\theta)]\text{  for  }z>0;\\
\mbf f_\mbf k(\mbf r)\cdot\mbf e_\nu(\mbf r)=e^{i(k_x-q_x)x-\kappa_x|x|}e^{(\kappa_1+\kappa_2)z}2\frac{\omega}{\omega_p}(\frac{\kappa_1}{\omega_p}-\sin\theta\frac{\omega^2}{\omega_p^2})\text{  for  }z<0.
\end{gather*}

Taking $r\approx1$ and retaining only the leading terms, we find
\begin{equation}
\int d\mbf r\varepsilon(\mbf r)\mbf f_\mbf k(\mbf r)\cdot\mbf e_\nu(\mbf r)\approx2\left(\sin\theta\frac{\kappa_1}{\kappa_1^2+k_z^2}-\frac1{\kappa+\kappa_2}\right)\frac{2\kappa_x}{\kappa_x^2+(k_x-q_x)^2}
\end{equation}
where we used $\int dxe^{i(k_x-q_x)x-\kappa_x|x|}=2\text{Re}\frac{1}{\kappa_x-i(k_x-q_x)}$ and $\int_0^{\infty} dze^{-(\kappa_1\pm ik_z)z}=\frac{1}{\kappa_1\pm ik_z}$. We can ignore the second term in the integration over $z$ since it is of order $1/\omega_p$ as opposed to $\omega_p/\omega^2$.

We now use the expressions $V_\omega=L_y/2\kappa_x\kappa_1$ and $\frac1V\sum_\mbf k=\frac1{(2\pi)^3}\int dk_xdk_y\int_0^{\infty}dk_z$ (where the $k_z$ wavevectors are restricted to positive values because of the halfspace occupied by the metal). We obtain
\begin{gather}
\gamma_{\omega,e}=32\pi\omega^2\kappa_x\kappa_1\int\frac{dk_xdk_z}{4\pi^2}\delta(\omega-\sqrt{k_x^2+k_z^2})\frac{\kappa_x^2}{\kappa_x^2+(k_x-q_x)^2}\frac{k_x^2}{\omega^2}\frac{\kappa_1^2}{(\kappa_1^2+k_z^2)^2}=\\
=\frac4\pi\omega^2\kappa_x^3\kappa_1^3\int_0^{\omega}dk_z\frac{\sqrt{\omega^2-k_z^2}}{\omega}\frac1{(\kappa_x^2+(q_x-\sqrt{\omega_2-k_z^2})^2)^2(\kappa_1^2+k_z^2)^2}=\\=\frac4\pi\omega\left(\frac{\kappa_x\kappa_1}{\omega^2}\right)^3\int_0^1dz\frac{\sqrt{1-z^2}}{((\kappa_x\omega)^2+(q_x/\omega-\sqrt{1-z^2})^2)^2((\kappa_1/\omega)^2+z^2)^2}
\end{gather}

Where $z=k_z/\omega$. The integral over $k_y$ simplifies with the integrals over $y$ and with $L_y$ arising from $V_\omega$. Since $z$ is limited by $\kappa_1/\omega=\omega/\omega_p\ll1$ in the last term, we can assume that most of the integral comes from small values of $z$ and thus approximate for $z\ll1$ and extend the integral to infinity.

\begin{gather}
\gamma_{\omega,e}=\frac4\pi\omega\left(\frac{\kappa_x\kappa_1}{\omega^2}\right)^3\int_0^\infty dz\frac{1}{(\frac{\kappa_x^2}{\omega^2}+(\frac{\omega^2}{2\omega_p^2}+\frac{z^2}2)^2)^2(\frac{\kappa_1^2}{\omega^2}+z^2)^2}=\\
\frac4\pi\omega\left(\frac{\kappa_x\kappa_1}{\omega^2}\right)^3\int_0^\infty dz\frac{16}{((\varepsilon'')^2\frac{\omega^8}{\omega_p^8}+(\frac{\omega^2}{\omega_p^2}+z^2)^2)^2(\frac{\omega^2}{\omega_p^2}+z^2)^2}\approx\\
\approx64\pi\omega\left(\frac{\kappa_x\kappa_1}{\omega^2}\right)^3\int_0^\infty dz\frac1{(\frac{\omega^2}{\omega_p^2}+z^2)^6}=8064\omega\frac{\varepsilon''^3}8\left(\frac{\omega^4}{\omega_p^4}\right)^3\left(\frac{\omega}{\omega_p}\right)^3\frac{\omega}{\omega_p}\left(\frac{\omega}{\omega_p}\right)^{-12}=\\
=1008\omega\varepsilon''^3\left(\frac{\omega}{\omega_p}\right)^4=1008\omega\left(\frac{\sigma(\omega)}{\epsilon_0\omega_p}\right)^3\left(\frac{\omega}{\omega_p}\right)
\end{gather}

\section{Solution of the Lindblad equation}\label{app:limblad_solution}
In this appendix we provide the steady state solution for the Lindblad equation in Eq.~\eqref{Eq:limblad_1mode}, dropping the index $\nu$ for simplicity. Since the unitary part is quadratic and the Lindbladian part consists of single particle losses and gains, the density matrix of each mode $\nu$ will be Gaussian. In order to find the steady state we use the Wigner function representation \cite{VogelRisken_pra89_wigner,wallsmilburn_quantumoptics}. For a bosonic mode described by a certain density matrix $\hat{\rho}$, this is generally defined as:
\begin{align}
    W(\alpha)=\int \frac{d^2\xi}{\pi^2} \;\chi(\hat{\rho};\xi) e^{\alpha\xi^* -\alpha^* \xi}
\end{align}
where $\chi$ is the characteristic function:
\begin{equation}
    \chi(\hat{\rho};\xi)= Tr\left[\hat{\rho}\,e^{\xi \hat{a}^\dagger-\xi^* \hat{a}}\right]\;.
\end{equation}
The Wigner function is normalized to 1 and is by definition a Gaussian for Gaussian states. In particular a thermal state for a mode at frequency $\omega$ at temperature $T$ reads:
\begin{equation}
    \hat{\rho}_T= \sum_{n=0}^\infty e^{-n\omega/T} \ket{n}\bra{n} \qquad W(\alpha)= \frac{1}{\pi\sigma_T^2}e^{-|\alpha|^2/\sigma_T^2}
\end{equation}
with $\sigma_T^2 =1/2+n_B(T/\omega)$. 
Following \cite{VogelRisken_pra89_wigner}, the time-evolution for the Wigner function in our case of two baths can be written as:
\begin{align}\label{Eq.wigner_evolution}
    \partial_t W(\alpha,t)= -\partial_{\alpha} \left[ \alpha (-i\omega -\frac{\Gamma^- -\Gamma^+}{2})W(\alpha,t) \right]-\partial_{\alpha^*} \left[ \alpha^* (i\omega -\frac{\Gamma^- -\Gamma^+}{2})W(\alpha,t) \right]+ \frac{\Gamma^- +\Gamma^+}{2}\partial_{\alpha}\partial_{\alpha^*} W(\alpha,t)
\end{align}
where $\Gamma^{\pm}=\gamma_e^{\pm}+\gamma_m^{\pm}$ is the total rate coming from a combination of the two baths. From this we can find the steady state by simply using a thermal state ansantz and setting $\partial_t W=0$. Note that the steady state has to be a thermal state as the Linblad equation has the same structure of that of a single bath. We find that there is a unique solution: 
\begin{align}
    W_{ss}(\alpha)=\frac{1}{\sigma_{T^*}^2\pi} \exp(-\frac{|\alpha|^2}{\sigma_{T^*}^2}) \qquad \qquad\sigma_{T^*}^2 =\frac{1}{2}+ \frac{\Gamma^+}{\Gamma^- -\Gamma^+}
\end{align}
The effective temperature $T^*$ can then be made explicit by identifying $n_B(T^*/\omega)\equiv \frac{\Gamma^+}{\Gamma^- -\Gamma^+}$, which leads to:
\begin{align}
    T^{*}= \frac{\omega}{-\log\left(r_{\nu,e}^- e^{-\omega/T_e} + r_{\nu,m}^- e^{-\omega /T_m} \right)}
\end{align}
as in Eq.~\ref{eq:Teff}. Note that for a single environment, for example $r_m^-=0$, this expression reduces to $T^*=T_e$. 

We can also look at the relaxation time-scales of the mode. For example the relaxation of a finite coherence can be derived from Eq. \eqref{Eq.wigner_evolution} and using $\langle\hat{a}\rangle_t= \int d^2\alpha\;  \alpha W(\alpha,t)$; giving:
\begin{equation}
    \frac{\partial}{\partial t}\langle \hat{a}\rangle = \left(-i\omega + \frac{\Gamma^-+\Gamma^+}{2}\right)\langle \hat{a}\rangle\;.
\end{equation}
This describes exponentially decaying oscillations with a rate $\Gamma^-+\Gamma^+=\gamma_e+\gamma_m$ as discussed in the main text. A similar rate also govern relaxation of other photon observables such as the photon number which controls the effective temperature.
\end{document}